\documentclass[aps,prx,twocolumn,showpacs,floatfix, superscriptaddress]{revtex4-2}

\usepackage{subfigure}
\usepackage{psfrag,graphicx}
\usepackage{pict2e}
\usepackage{dcolumn}
\usepackage{amsmath,amssymb,braket}
\usepackage{bm}
\usepackage{color}
\usepackage{latexsym}
\usepackage{epstopdf}
\usepackage{color}
\usepackage{comment}
\usepackage[english]{babel}
\UseRawInputEncoding
\usepackage{soul}
\usepackage{amsfonts}
\usepackage{bm}
\usepackage{natbib}
\usepackage{bbold}

\usepackage{enumerate}

\usepackage{tikz}

\definecolor{myblue}{rgb}{0.2,0.2,0.8}
\definecolor{myzard}{cmyk}{0,0,0.05,0}
\definecolor{mywhite}{rgb}{1,1,1}
\definecolor{mywhite}{rgb}{1,1,1}
\definecolor{myred}{rgb}{1,0.,0.3}
\usepackage[colorlinks=true,citecolor=myblue,linkcolor=myred]{hyperref}

\definecolor{darkgreen}{rgb}{0.0, 0.4, 0.26}

\definecolor{mygrey}{gray}{0.35}
\definecolor{myblue}{rgb}{0.2,0.2,0.8}
\definecolor{myzard}{cmyk}{0,0,0.05,0}
\definecolor{mywhite}{rgb}{1,1,1}
\definecolor{mywhite}{rgb}{1,1,1}
\definecolor{myred}{rgb}{1,0.,0.3}

\def\be{\begin{equation}}
\def\ee{\end{equation}}
\def\ba{\begin{align}}
\def\enda{\end{align}}
\def\bi{\begin{itemize}}
\def\ei{\end{itemize}}

\def\beq{\begin{equation}}
\def\beq{\begin{equation}}
\def\eeq{\end{equation}}

\begin{document}

\title{Photon-mediated interactions by Floquet photonic lattices}

 \author{Jia-Qiang Chen}
 \affiliation{Ministry of Education Key Laboratory for Nonequilibrium Synthesis and Modulation of Condensed Matter,
 Shaanxi Province Key Laboratory of Quantum Information and Quantum Optoelectronic Devices,  School of Physics, Xi’an Jiaotong University, Xi’an 710049, China}
  \affiliation{Institute of Fundamental Physics IFF-CSIC, Calle Serrano 113b, 28006 Madrid, Spain}
\author{Peng-Bo Li}
\email{lipengbo@mail.xjtu.edu.cn}
 \affiliation{Ministry of Education Key Laboratory for Nonequilibrium Synthesis and Modulation of Condensed Matter,
 Shaanxi Province Key Laboratory of Quantum Information and Quantum Optoelectronic Devices,  School of Physics, Xi’an Jiaotong University, Xi’an 710049, China}
  \author{\'Alvaro G\'omez-Le\'on}
     \email{a.gomez.leon@csic.es}
   \affiliation{Institute of Fundamental Physics IFF-CSIC, Calle Serrano 113b, 28006 Madrid, Spain}
\author{Alejandro Gonz\'alez-Tudela}
\email{a.gonzalez.tudela@csic.es}
  \affiliation{Institute of Fundamental Physics IFF-CSIC, Calle Serrano 113b, 28006 Madrid, Spain}

\begin{abstract}
We investigate the interactions between two-level emitters mediated by time-dependent, one-dimensional, structured photonic baths, focusing on Floquet topological lattices. Building on the framework of periodically driven photonic lattices, we demonstrate and characterize the emergence of tunable-range emitter's interactions mediated by bound states absent in static photonic lattices. In particular, we show that one can not only obtain different spatial interaction dependencies with respect to the static bath scenarios, but also in qualitatively different regimes due to the time-dependent nature of the bath, for example, when the emitters have different frequencies. This work sheds light on the interplay between non-equilibrium photonics and quantum optics and can serve as the basis for analyzing Floquet photonic lattices in higher dimensions.
\end{abstract}

\maketitle


\section{Introduction}~\label{sec:intro}

Engineering versatile photon-mediated interactions between quantum emitters (QE) can be instrumental for quantum simulation~\cite{Gonzalez-Tudela2015b,douglas15a,Armon2021,Tabares2023VariationalQED}, metrology~\cite{Bornet2023,Eckner2023}, and information technologies (see Refs.~\cite{Chang2018, Sheremet2023WaveguideCorrelations,Gonzalez-Tudela2024} for recent reviews on the topic). Structured photonic lattices, like photonic crystals~\cite{joannopoulos97a} or coupled-cavity arrays resonators, provide a way to achieve it due to their potential to engineer complex band-structures, translating into non-trivial photon-mediated interactions. This is motivating a widespread theoretical~\cite{calajo16a,Shi2016,Shi2018,Gonzalez-Tudela2018,Calajo2019b,Gonzalez-Tudela2019a,Bello2019a,Leonforte2020b,deBernardis2021,Perczel2020a,Perczel2020b,Vega2021a,rocatti2022,Vega2023TopologicalQED,Garcia-Elcano2020,Garcia-Elcano2021,GarciaElcano2023,rocatti2024,Bello2023,Dibenedetto2024,Leonforte2024,Du2024,Salinas2024HarnessingLattice,Vicenzio2025,Lanuza2022,Lanuza2024} and experimental~\cite{goban13a,thompson13a,goban15a,hood16a,Beguin2020a,Zhou2023,Kim2019a,Samutpraphoot2020,Dordevic2021,Menon2024,Tiranov2023,evans18a,Lukin2022a,Rugar2020,Rugar2021,liu17a,Sundaresan2019,Mirhosseini2019,Scigliuzzo2022,Kim2020b,Zhang2022,Owens2022,krinner18a,Kwon2022FormationLattice,Kim2024a,CastilloMoreno2024} interest in studying the photon-mediated interactions induced by these structured photonic lattices. A particularly noteworthy example is the case of topological photonic lattices~\cite{Bello2019a,Leonforte2020b,deBernardis2021,Perczel2020a,rocatti2022,Perczel2020b,Vega2021a,Vega2023TopologicalQED,Garcia-Elcano2020,Garcia-Elcano2021,GarciaElcano2023,rocatti2024}, where the interactions are mediated by the appearance of qubit-photon bound states linked to the topological edge modes of the lattice. In all these works, however, the photonic lattice properties are fixed while mediating the interactions, which limits the versatility of the interactions that can be obtained.

A way of gaining versatility is considering time-dependent photonic media, a topic of rising interest in the photonic community~\cite{Galiffi2024}. In these systems, the photonic properties of the media, like refractive index, are modulated in space and time to achieve unique phenomena, such as non-reciprocitiy~\cite{Sounas2017} or  amplification~\cite{lustig2018}. Among these phenomena, are the emergence of Floquet topological phases~\cite{Oka2019,Lidner2011,Grushin2014,DiazFernandez2019,Qin2022} appearing in photonic lattices subject to periodic temporal modulations. 
These phases can host topological edge states without an equilibrium counterpart. For example, it is possible to have topologically protected edge states coexisting with trivial Floquet bands, referred to as Anomalous Floquet phases~\cite{Kitagawa2010,Rudner2013,Nathan2015,Roy2017,GomezLeon2024,Gavensky2024}, as experimentally confirmed in several platforms~\cite{Wang2013a,Wintersper2020,Adiyatullin2022,ElSokhen2024}. Thus, a natural question is whether the edge states in these phases can also be used to mediate interactions between emitters, and what are their distinctive features compared to their static counterparts~\cite{Bello2019a,Leonforte2020b,deBernardis2021,Perczel2020a,rocatti2022,Perczel2020b,Vega2021a,Vega2023TopologicalQED,Garcia-Elcano2020,Garcia-Elcano2021,GarciaElcano2023,rocatti2024}.

\begin{figure}[tb]
    \centering
\includegraphics[width=0.85\linewidth]{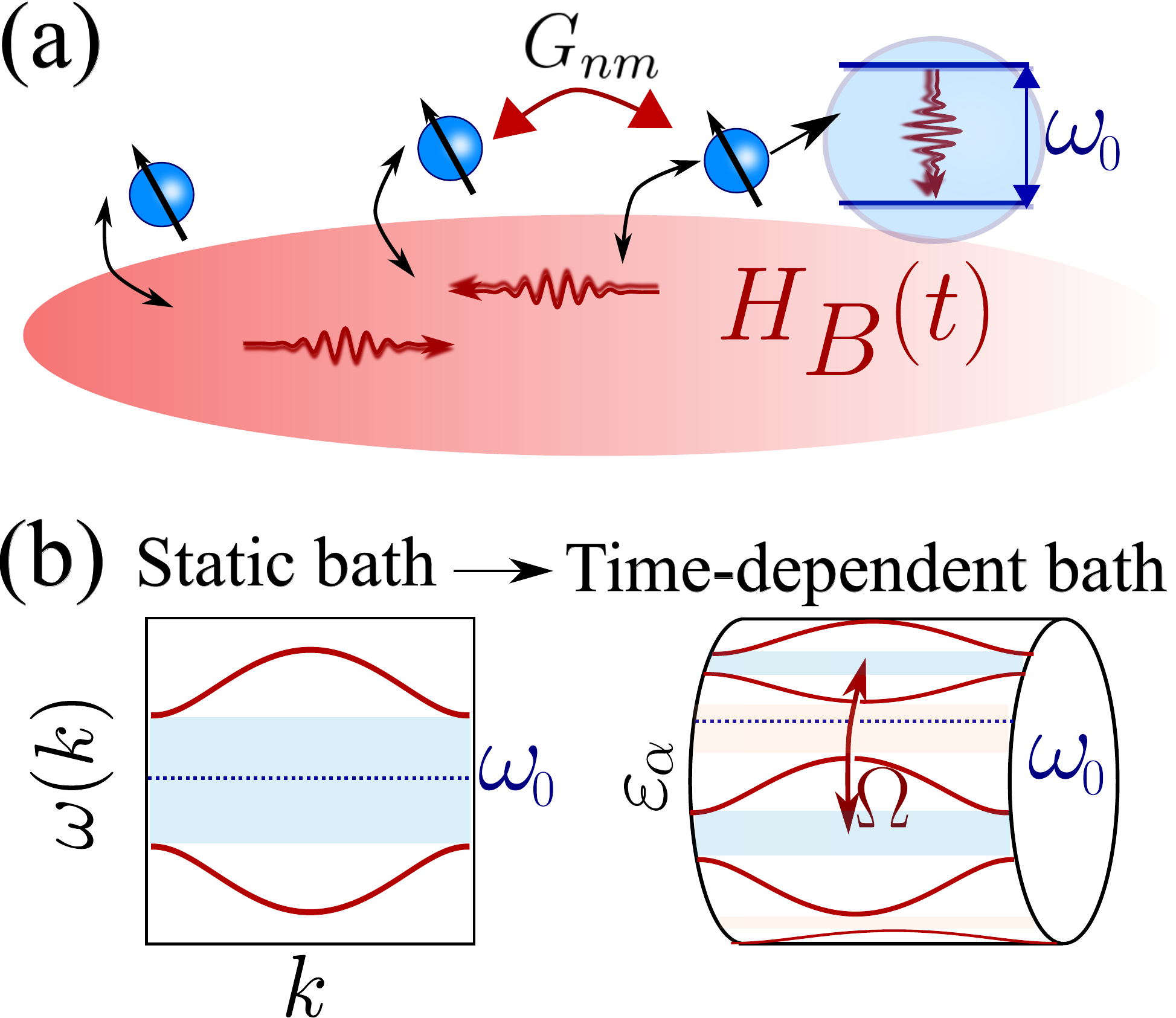}
    \caption{(a) Schematic representation of the setup: two-level emitters, with frequency $\omega_0$, are locally coupled to a photonic bath described by a time-dependent, periodic Hamiltonian $\hat{H}_B(t)$ with a characteristic frequency $\Omega$. When the emitter's frequency lies in a photonic band-gap, this coupling results in coherent emitter interactions, $G_{nm}$, whose shape is determined by the photonic band-structure. (b) Comparison between the static and dynamic photonic band-structures. In the static case, the photonic bands $\omega(k)$ and band-gaps (in shaded blue) are determined by the geometry. In the time-dependent case, on top of these geometrical band-gaps, new temporally induced band-gaps appear (shaded orange) in the quasi-energy energy structures $\varepsilon_\alpha$. These are denoted as $0$ and $\pi$-gaps, respectively}
    \label{fig:0}
\end{figure}

Here, we provide an answer to this question by studying the interactions arising from one-dimensional photonic lattices with periodically modulated hopping~\cite{GomezLeon2024} (see Fig.~\ref{fig:0}(a)).
Using exact numerical calculations, we demonstrate the emergence of coherent photon-mediated exchanges between emitters due to the appearance of out-of-equilibrium emitter-photon bound states~\cite{bykov75a,john90a,kurizki90a}. In contrast to the static case, the quasienergy spectrum can display topological bound states not only in the $0$-gap, but also in the $\pi$-gap, which is the one separating different Floquet sidebands (see Fig.\ref{fig:0}(b) for a description of the two nonequivalent gaps), introducing an additional frequency window to mediate interactions between emitters. Moving to an interaction picture of the light-matter Hamiltonian and by neglecting the fast-rotating terms, we also obtain semi-analytical approximations for the shape of these bound states and their effective coherent interactions. Remarkably, we find that coherent interactions can be obtained in frequency regions that would be highly dissipative for the static photonic lattice, and also in situations where the frequencies of the emitters are different, something not possible with static photonic baths. These observations demonstrate the potential of time modulations to change both the nature and the shape of the photon-mediated interactions between emitters.

The manuscript is structured as follows: in Section~\ref{sec:setup}, we introduce the general setup that we consider in this manuscript, writing both the light-matter Hamiltonian and analyzing the emergence of anomalous phases in the photonic part. Next, in Section~\ref{sec:interactions}, we analyze the emergence of coherence excitation exchanges by studying both the single and many emitters' regimes. Finally, we summarize our findings in Section~\ref{sec:conclusion}.

\section{Setup~\label{sec:setup}}

\begin{figure}[tb]
    \centering
    \includegraphics[width=0.95\linewidth]{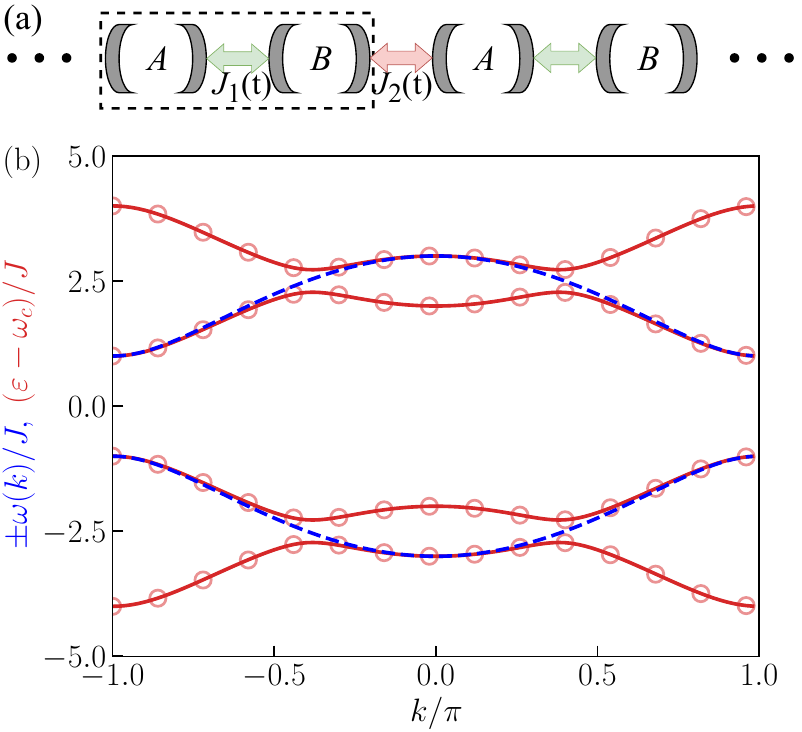}
    \caption{(a) Schematic of the time-dependent coupled-cavity array $\hat{H}_B(t)$ used in this manuscript. It consist in a dimerized photonic lattice with time-dependent intra(inter) dimer hopping $J_{1(2)}(t)$. (b) Comparison between the energy dispersion of the static model  of Eq.~\eqref{eq:static} (blue dashed line) and the exact quasi-energy of dynamical lattice (red-solid line) calculated from Eq.~\eqref{eq:quasienergies}. For the parameters chosen ($V=0.2J$, $\Omega=5J$, and $J'=2J$), the exact solution of the quasi-energy perfectly matches the analytical solution of Eq.~\eqref{eq:varepsik} (red circle markers) obtained by neglecting fast rotating terms of the bath.}
    \label{fig:model}
\end{figure}

Here, we review in~\ref{subsec: photonic bath} the properties of the one-dimensional photonic lattice that we consider to be exchanging excitations with the emitter. In particular, we identify the parameter regimes where novel, time-dependent, topological edge modes appear, and compare their properties with the standard static case of one-dimensional topological photonic lattices. Next, in section~\ref{subsec:lmHam}, we write the full light-matter interaction Hamiltonian including the emitter degrees of freedom that we use for our exact numerical simulations.

\subsection{Static vs Floquet topological edge states~\label{subsec: photonic bath}}

Along this manuscript, we consider the photonic lattice model depicted in Fig.~\ref{fig:model}(a). It consists in a coupled-cavity array with a bipartite lattice structure and temporally modulated hoppings. It is a time-dependent version of the Su-Schrieffer-Heeger (SSH) model~\cite{Su1979}, which already features topological edge states in its static version. The Hamiltonian describing the photonic lattice $\hat{H}_B(t)=\hat{H}_{\mathrm{SSH}} + \hat{H}_d(t)$ consists of a time-independent part given by (setting $\hbar=1$ for the rest of the manuscript):
\begin{align}
\hat{H}_{\mathrm{SSH}} =& \omega_c \sum_{j=1}^N  \left( \hat{a}_j^\dagger \hat{a}_j + \hat{b}_j^\dagger \hat{b}_j \right)
\\
&+ \sum_{j=1}^N \left( J \hat{a}_j^\dagger \hat{b}_j + J' \hat{a}_{j+1}^\dagger \hat{b}_j  + \mathrm{H.c.} \right)\nonumber,
\end{align}
and a time-modulated hopping term given by:
\begin{align}
\hat{H}_d(t) = 2V \cos(\Omega t) \sum_j \left( \hat{a}_j^\dagger \hat{b}_j - \hat{a}_{j+1}^\dagger \hat{b}_j + \mathrm{H.c.} \right),\label{eq:timeH}
\end{align}
where \(\hat{a}_j^\dagger\) (\(\hat{a}_j\)) and \(\hat{b}_j^\dagger\) (\(\hat{b}_j\)) are the creation (annihilation) operators for the two sublattices $A/B$ at site \(j\), whose sum run over the number of unit cells of the lattice, denoted by $N$.
We assume that the $A/B$ modes have the same energy $\omega_c$.
\(J\) and \(J'\) represent the static components of the intra-cell and inter-cell hoppings, respectively,  and \({V,\Omega}\) represent modulation amplitude and frequency of the driven hoppings, respectively.
Therefore, the total time-dependent intra-cell and inter-cell hoppings are:
\begin{align}
J_1(t) &= J + 2V \cos(\Omega t), \\
J_2(t) &= J'- 2V \cos(\Omega t),
\end{align}

For the case of periodic boundary conditions (PBC), we can define the lattice operators in momentum space:
\begin{align}
\hat{a}_k [\hat{b}_k] &= \frac{1}{\sqrt{N}} \sum_{j=1}^N \hat{a}_j [\hat{b}_j] e^{-ikj},
\end{align}
and with these definitions, we can rewrite the bath Hamiltonian as:
\begin{align} \label{eq:H_k}
\hat{H}_B(t)=&\sum_k \left[\omega_c \hat{a}_k^\dagger \hat{a}_k +\omega_c\hat{b}_k^\dagger \hat{b}_k + \left(h_k \hat{a}_k^\dagger \hat{b}_k +  \mathrm{H.c.}\right)\right]  \notag \\
    &+\sum_k\left( 2d_k  \cos(\Omega t) \hat{a}_k^\dagger \hat{b}_k + \mathrm{H.c.}\right),
\end{align}
where $h_k=J+J'e^{-ik}=\omega(k) e^{i\theta_k}$ and $d_k=V(1-e^{-ik})=|d_k| e^{i\beta_k}$ are the time-independent and time-dependent hopping amplitudes, respectively.

In the static situation, i.e., $V = 0$, the photonic Hamiltonian $\hat{H}_{\mathrm{SSH}}$ can be diagonalized by the following the operators:
\begin{align}
\hat{u}_{k} [\hat{l}_{k}] = \frac{1}{\sqrt{2}} \left( \hat{a}_k +[-] e^{i\theta_k} \hat{b}_k \right) ,
\end{align}
With these operators, the photonic Hamiltonian reads:
\begin{equation}
\hat{H}_{\text{SSH}} = \sum_k\left[ (\omega_c+\omega(k)) \hat{u}_{k}^\dagger \hat{u}_{k} +(\omega_c-\omega(k)) \hat{l}_{k}^\dagger \hat{l}_{k} \right],\label{eq:static}
\end{equation}
where $\omega(k) = \sqrt{J^2+(J')^2+2 J J' \cos(k)}$. We can see that the band structure in the static case results in two bands with energies $\omega_c\pm \omega(k)$ (see dashed blue lines in Fig.~\ref{fig:model}(b)). However, when $J = J'$, the two bands touch at $k=\pm \pi$, and we recover the limit of a coupled-cavity array with uniform, nearest-neighbour hoppings. Importantly, when $J' > J$, the band-gap hosts topological zero-energy modes in finite systems~\cite{asboth15}.

\begin{figure}[tb]
    \centering
    \includegraphics[width=0.95\linewidth]{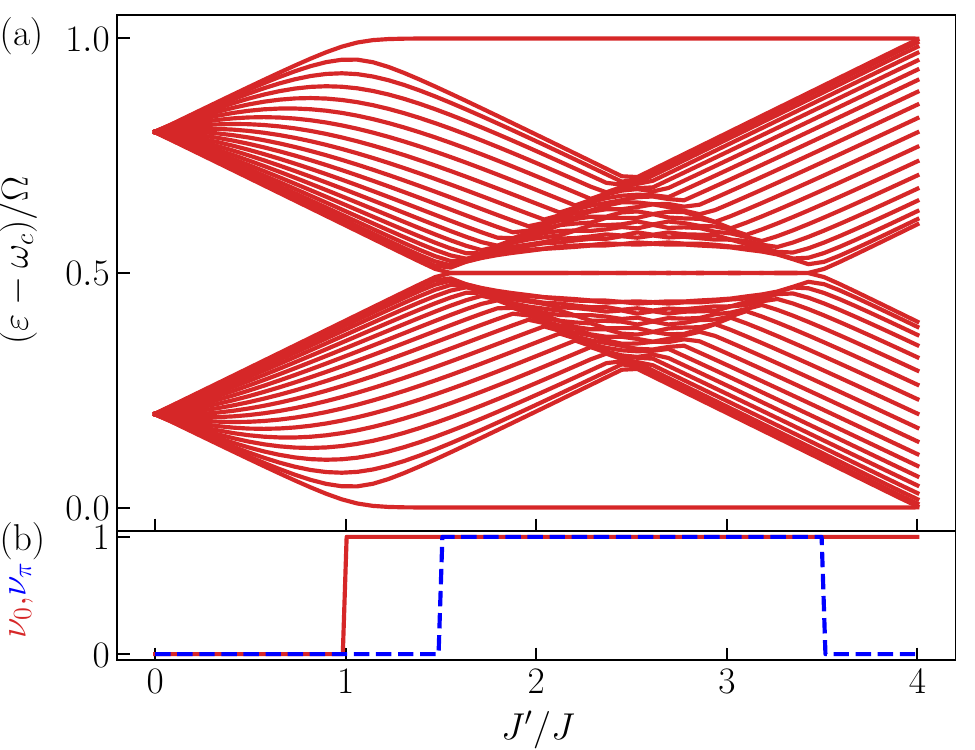}
    \caption{(a) Quasi-energy spectrum of finite coupled-cavity array with $N=20$ unit cells as a function of the ratio $J'/J$. (b) Winding number associated to the $0$- and $\pi$-gaps defined in Eqs.~\eqref{eq:winding0} and~\eqref{eq:winding_pi}, respectively. These winding numbers evidence that the $0$-gap closes at $J'=J$ and the $\pi$-gap at $\Omega=2|J\pm J'|$, which are consistent with the quasi-energy spectrum of panel (a). Parameters: $V=0.2J$, $\Omega=5J$.}
    \label{fig:quasienergy_OBC}
\end{figure}

In the time-modulated situation, i.e., $V \neq 0$, the photonic Hamiltonian $\hat{H}_B(t)$ cannot be directly diagonalized using the static operators due to its explicit time-dependence.
Since the bath possesses the periodicity $\hat{H}_B(t + T) = \hat{H}_B(t)$ with $T = 2\pi / \Omega$, the Floquet formalism allows us to write the solutions of the time-dependent Schr$\ddot{\text{o}}$dinger equation in terms of time-dependent states of the form:
\begin{equation}
|\Psi_\alpha(t)\rangle = e^{-i \varepsilon_\alpha t} |\Phi_\alpha(t)\rangle,
\end{equation}
where $|\Phi_\alpha(t)\rangle$ is a Floquet state with the same period as the Hamiltonian, $|\Phi_\alpha(t + T)\rangle = |\Phi_\alpha(t)\rangle$, and $\varepsilon_\alpha$ is its corresponding quasi-energy.
The Floquet quasi-energy spectrum is determined from the eigenvalue problem:
\begin{equation}
\hat{U}(t_0+T,t_0) |\Phi_\alpha(t_0)\rangle = e^{-i\varepsilon_\alpha T} |\Phi_\alpha(t_0)\rangle\,,\label{eq:quasienergies}
\end{equation}
where $\hat{U}(t_1,t_0)=\hat{\mathcal{T}}e^{-i\int_{t_0}^{t_1}  \hat{H}_B(\tau)\,d\tau}$ is the time evolution operator from time $t_0$ to time $t_1$ and $\hat{\mathcal{T}}$ is the time-ordering operator. Based on this, we can calculate the quasi-energy spectrum, which is composed of an infinite number of sidebands separated by a period $\Omega$, due to the fact that the phases fulfill: $e^{-i\varepsilon_\alpha T}=e^{-i(\varepsilon_\alpha T+2n\pi)}$.
We refer to the quasi-energy range $[\omega_c-\Omega/2,\omega_c+\Omega/2)$ as the first "Floquet-Brillouin zone".
In Fig.~\ref{fig:model}(b) we show, for $\Omega/J=5$, the zero sideband anti-crossing with the neighboring ones, as a function of $k$ (red-solid line).
We label $\pi$-gap the one separating different sidebands and $0$-gap to the one separating the bands within the same Floquet-Brillouin zone.
In  finite systems with open boundary conditions (OBC), both gaps can host topological edge states.

Numerous approximations and methods have been proposed to analyze Floquet systems in different frequency regimes~\cite{Goldman2015,Perez2015,Nathan2015,Rudner2020,Rodriguez2018,Vogl2020,GomezLeon2024}.
Here, we use the formalism developed in Ref.~\cite{GomezLeon2024}, especially useful to deal with the intermediate-frequency regime $\Omega/2\in\omega(k)$ which we are interested in, and where $0$-gap and $\pi$-gap edge states can coexist. This formalism performs first a transformation to the interaction picture with respect to the driven part of the hopping, $\hat{R}_d(t)=e^{-i\int \hat{H}_d(t) \, dt}$, to be able to capture strong field effects. This transformation transfers the time-dependence from $H_d(t)$ into a hopping with a non-linear time-dependent phase. Applying the well-known Jacobi-Anger expansion:
\begin{align}
    \cos\left[z\sin(\theta)\right] 
    &= \mathcal{J}_0(z) + 2\sum_{m=1}^\infty \mathcal{J}_{2m}(z) \cos(2m\theta), \\ 
    \sin\left[z\sin(\theta)\right] 
    &=  2\sum_{m=1}^\infty \mathcal{J}_{2m-1}(z) \sin\left[(2m-1)\theta\right],
\end{align}
where $\mathcal{J}_m(z)$ is the $m$-th Bessel function of the first kind, we obtain the following approximation of the time-dependent Hamiltonian (see Appendix~\ref{App:effective Hamiltonian}):
\begin{align}
    \hat{H}_B^d(t) &\approx \sum_k \left[\omega_c\left(\hat{a}_k^\dagger \hat{a}_k + \hat{b}_k^\dagger \hat{b}_k \right) + \left(h_k^d \hat{a}_k^\dagger \hat{b}_k + \mathrm{H.c.}\right)\right] \notag  \\ 
    &+\sum_k 2\gamma_k \sin(\Omega t) \left(\hat{a}_k^\dagger \hat{a}_k - \hat{b}_k^\dagger \hat{b}_k\right),\label{eq:trans2}
\end{align}
where, for practical purposes, we have kept only the terms with $\mathcal{J}_0(z)$ and $\mathcal{J}_1(z)$, and we have defined:
\begin{align}
    h_k^d &= h_k\frac{1+\mathcal{J}_0(2z_k)}{2} + e^{2i\beta_k}h_k^* \frac{1-\mathcal{J}_0(2z_k)}{2}=\omega_d(k)e^{i\theta_k^d}, \notag \\
   \gamma_k &=\mathcal{J}_1(2z_k) (J+J')V\sin(k)/|d_k|,
\end{align}
with $z_k=2|d_k|/\Omega$. Notice that $h_k^d$ and $\gamma_k$ are the renormalized hopping and energy, respectively, due to the dressing of the driving field.
Importantly, it is expected that the truncation to only the zeroth and the first Bessel functions works well when one-photon resonances dominate over two-photons resonances.

The first line in Eq.~\eqref{eq:trans2} is time-independent and defines the renormalized bands due to the periodic drive. Then, it can be diagonalized as in the static case, by the operators: $\hat{u}_{k} [\hat{l}_{k}] = \frac{1}{\sqrt{2}} \left( \hat{a}_k +[-] e^{i\theta_k^d} \hat{b}_k \right)$, just changing the phase to $\theta_k^d$ to account for the renormalized hopping terms. In this new basis the Hamiltonian for the bath reads:
\begin{align}
    \hat{H}_B^d(t) &=\sum_k \left[(\omega_c+\omega_d(k))\hat{u}_k^\dagger \hat{u}_k + (\omega_c-\omega_d(k))\hat{l}_k^\dagger \hat{l}_k\right] \notag \\ 
    &+ \sum_k 2\gamma_k \sin(\Omega t) \left(\hat{u}_k^\dagger \hat{l}_k + \hat{l}_k^\dagger \hat{u}_k\right).  \label{eq:trans3}
\end{align}
Importantly, this Hamiltonian remains valid for strong drive due to the Bessel function renormalization. Similarly to the static case, the time-independent part has two bands with energies $\omega_c\pm \omega_d(k)$. However, now the time modulation produces their hybridization. To study this hybridization in more detail, we can apply a rotating wave approximated (RWA) to simplify Eq.~\eqref{eq:trans3}. This requires to perform a transformation to the interaction picture, using the time-independent part of Eq.~\eqref{eq:trans3}, and remove fast rotating terms. Finally, going to a rotating frame with the frequency of the drive, $\hat{u}_k \to \hat{u}_k e^{-i \frac{\Omega}{2} t}$ and $\hat{l}_k \to \hat{l}_k e^{i \frac{\Omega}{2} t}$, we obtain a time-independent, RWA Hamiltonian which reads:
\begin{align}
    \hat{H}_B^\mathrm{RWA} &\approx \sum_{k}\Bigg\{\left(\omega_c +\omega_d(k)-\frac{\Omega}{2}\right)\hat{u}_k^\dagger \hat{u}_k \label{eq:HBRWA1} \\ 
    &+\left(\omega_c -\omega_d(k)+\frac{\Omega}{2}\right) \hat{l}_k^\dagger \hat{l}_k + i\gamma_k \hat{u}_k^\dagger \hat{l}_k - i\gamma_k \hat{l}_k^\dagger \hat{u}_k \Bigg\}\,\notag.
\end{align}

The advantage of this Hamiltonian $\hat{H}_B^\mathrm{RWA}$ is that, being time-independent, it can be easily diagonalized, yielding
\begin{align}
    \hat{H}_B^{\text{RWA}}=\sum_k \left[\omega_c+\lambda(k)\right]\hat{p}_k^\dagger \hat{p}_k + \left[\omega_c-\lambda(k)\right] \hat{q}_k^\dagger \hat{q}_k\,,\label{eq:HBRWA}
\end{align}
with eigenvalues $\lambda(k)=\sqrt{\left[\omega_d(k)-\Omega/2\right]^2+\gamma_k^2}$ and eigenoperators:
\begin{align}
    \hat{p}_k=\cos \left( \frac{\phi_k}{2}\right) \hat{u}_k + i\sin\left(\frac{\phi_k}{2}\right)\hat{l}_k, \\
    \hat{q}_k=\sin\left(\frac{\phi_k}{2}\right) \hat{u}_k - i \cos\left(\frac{\phi_k}{2}\right)\hat{l}_k,
\end{align}
where $\phi_k=\text{atan}\left[ \gamma_k,\, \omega_d(k)-\Omega/2\right]$.

By undoing the transformations previously used, we can go back to the laboratory frame and impose the time periodicity to find the Floquet states. Like this, we can finally determine the relationship between the quasi-energies and the eigenvalues which reads:
\begin{equation}
    \label{eq:varepsik}
    \varepsilon_{\pm}(k)=\omega_c+\frac{2m+1}{2}\Omega\pm\lambda(k)\,.
\end{equation}

From these expressions, it is clear the emergence of new band-gaps, when $V\neq 0$, due to the hybridization of the different sidebands via the driving field.  In Fig.~\ref{fig:model}(b), we compare the result of this analytical solution (in red circles) against the result of the exact quasi-energies (red solid line), showing an excellent agreement between them, for our choice of parameters.

Now, let us review what happens with these static and driving-induced band-gaps for finite systems with OBC. For that, we plot in Fig.~\ref{fig:quasienergy_OBC}(a) the quasi-energy spectrum, centered at the $\pi$-gap, for a finite system with OBC.
We observe that when $J'>J$, the $0$-gap hosts a pair of degenerate zero-energy modes (notice that the states at zero and $\Omega$ are formally the same). The topology in 0-gap is captured by the following winding number:
\begin{align}
    \nu_0=\frac{1}{2\pi}\int_{-\pi}^\pi \frac{\partial \theta_k^d}{\partial k}\, dk\,, \label{eq:winding0}
\end{align}
which characterizes the topology of the  stroboscopic part of the Hamiltonian of Eq.~\eqref{eq:trans2}. In Fig.~\ref{fig:quasienergy_OBC}(b), we plot this winding number in solid red, showing how it coincides with the emergence of the topological edge states in the $0$-gap of the quasi-energy spectrum of Fig.~\ref{fig:quasienergy_OBC}(a). If the system is in the weak driving regime, $V/\Omega\ll 1$, then $\theta_k^d\approx \theta_k$, and the topological phases coincide with the static case.

When the frequency $\Omega$ is in the range $2|J-J'|<\Omega<2|J+J'|$, the bath becomes resonant with the drive. In this situation, the $\pi$-gap closes and re-opens, hosting two additional edge states which have no counterpart in the equilibrium case. The reason is that periodically driven systems, due to the periodicity of the quasienergies, display a $\pi$-gap, which separates different Floquet copies. The edge states in this gap, as in the case of the $0$-gap, can be shown to be of topological origin by means of a winding number, $\nu_\pi$, which can be calculated from the Hamiltonian $\hat{H}_B^\text{RWA}$ of Eq.~\eqref{eq:HBRWA1} as follows:
\begin{equation}
    \nu_\pi = \frac{1}{4\pi}\int_{-\pi}^\pi \text{tr}\left\{\tau_\text{c} (\hat{H}_B^\text{RWA})^{-1} i \partial_k \hat{H}_B^\text{RWA}\right\} dk ,\label{eq:winding_pi}
\end{equation}
where $\tau_\text{c}$ is the chiral operator of the rotating frame Hamiltonian, $\{ \tau_\text{c}, \hat{H}_B^\text{RWA} \}=0$. 
Notice that $\hat{H}_B^\text{RWA}$ captures the topological invariant of the $\pi$-gap only, because the topological information about the $0$-gap invariant is encoded in the eigenoperators used to diagonalize the static part.
In Fig.~\ref{fig:quasienergy_OBC}(b), we plot in blue this winding number as a function of $J'/J$, showing how the region where $\nu_\pi=1$ coincides with the emergence of edge states in the $\pi$-gap in Fig.~\ref{fig:quasienergy_OBC}(a). The regions where the two topological invariants are non-zero, and thus the $0$ and $\pi$-gap edge states coexist in the quasispectrum, are labelled as the anomalous Floquet phases~\cite{Kitagawa2010,Rudner2013,Nathan2015,Roy2017,GomezLeon2024,Gavensky2024}. 

\begin{figure}[tb]
    \centering
   \includegraphics[width=0.95  \linewidth]{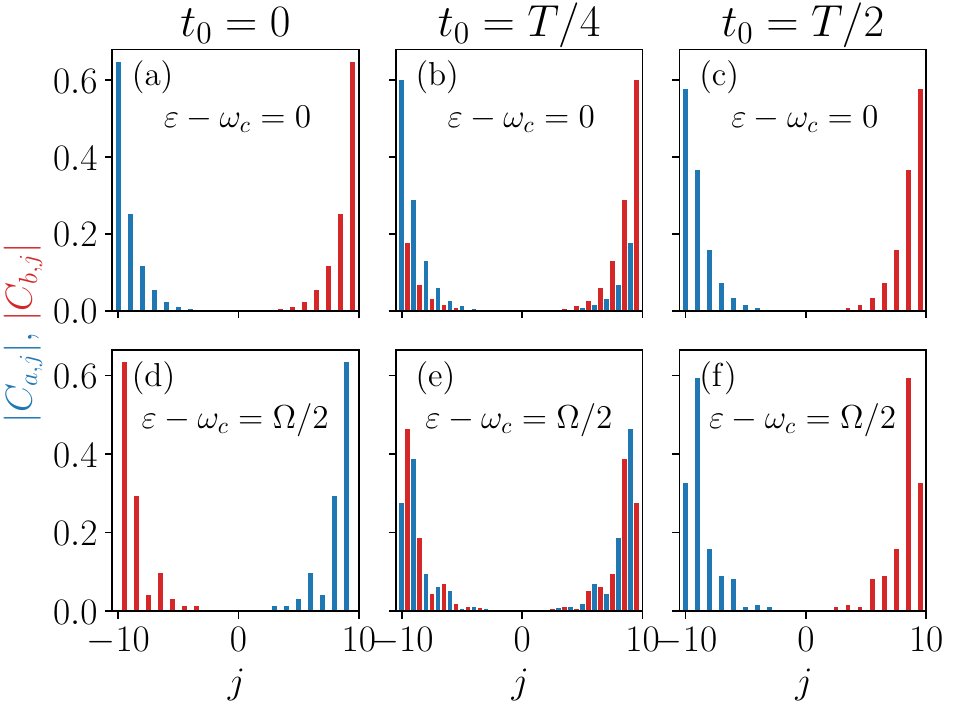}
    \caption{Spatio-temporal shape of the edge modes in $0$-gap (top row) and $\pi$-gap (bottom row) for a lattice with $J'=2J$, $V=0.4J$, $\Omega=5J$, $N=20$. Probability amplitudes $|C_{a,j}|$ are shown in blue, while the $|C_{b,j}|$ are shown in red.}
    \label{fig:edge_state}
\end{figure}

Next, we review the spatial shape of the static and time-dependent edge states to understand the similarities and differences between them. In both cases, the edge states can be written as a superposition of photons in the different sites, i.e., $\ket{\Phi_\mathrm{Edge}}=\sum_j\left(C_{a,j}\hat{a}_j^\dagger + C_{b,j}\hat{b}_j^\dagger\right)\ket{\mathrm{vac}}_B$, with $C_{\alpha,j}$ being the weight of the states in the $\alpha$-sublattice of the $j$-th unit cell.
The first thing to notice is that the bound-states appearing in these time-dependent media are, obviously, time-dependent, with a period $T=2\pi/\Omega$ determined by the frequency of the drive, $\Omega$. However, its spatial distribution strongly depends on the type of gap where they appear. To evidence that, in Fig.~\ref{fig:edge_state} we plot the spatial distribution of the edge states at the $0$- (upper row) and $\pi$-gap (lower rows) and for different times $t_0=0,T/4,T/2$ in the different columns. There, we see that the spatial localization of the $0$-gap edge states is very similar to the static scenario during the whole period: they are exponentially localized at the edges predominantly occupying a particular sublattice at each side during the time evolution.
On the contrary, the $\pi$-gap edge states display nonmonotonic localization, and the excitation of one side can reverse from $A(B)$ to $B(A)$ in half period.

\subsection{Light-matter interaction Hamiltonian~\label{subsec:lmHam}}

After reviewing the bath properties, let us now introduce the additional Hamiltonian that needs to be introduce to describe a situation like the one depicted in Fig.~\ref{fig:0}(a), i.e., when several emitters are locally coupled to the time-dependent photonic lattice. We choose a two-level system description for the emitters, that is, we assume they only have a ground ($g$) and excited ($e$) state level with dipole transition frequency $\omega_n$ for the $n$-th emitter. Thus, the Hamiltonian describing the emitters' dynamics is:
\begin{align}
    \hat{H}_S= \sum_n^{N_e}\omega_n \hat{\sigma}_n^\dagger\hat{\sigma}_n\,,
\end{align}
where $\hat{\sigma}_n =\ket{g}_n\bra{e}$ is the $n$-th emitter's dipole operator and $N_e$ is the emitter number. 

We also assume the emitters are locally interacting with the photonic lattice such that the light-matter interaction Hamiltonian is given by:
\begin{equation}
    \hat{H}_\mathrm{int} = \sum_{n} (\hat{\sigma}_n + \hat{\sigma}_n^\dagger)\left(g_{n,\alpha} \hat{\alpha}_{j_n}  + g_{n,\alpha}^* \hat{\alpha}_{j_n}^\dagger \right),
\end{equation}
where $\alpha=a,b$ is the index representing the sublattice the emitter couples to, $\hat{\alpha}_{j_n}^\dagger$ ($\hat{\alpha}_{j_n}$) is the creation (annihilation) operator for the photonic lattice at site $j_n$ and we assume the equal coupling strength $g_{n,\alpha}\equiv g$ between the $n$-th emitter and the $\alpha$-photonic mode at site $j_n$.
Now, by combining these three parts, we obtain the complete light-matter interaction Hamiltonian:
\begin{equation}
    \hat{H}(t) = \hat{H}_S + \hat{H}_B(t) + \hat{H}_\text{int}.
\end{equation}
This time-dependent Hamiltonian will serve as the basis for analyzing the emergent interactions induced by the interplay between the emitters and the temporally modulated photonic bath.

\section{Emitter-photon bound-state and interactions~\label{sec:interactions}}

\begin{figure}[tb]
    \centering
   \includegraphics[width=0.95\linewidth]{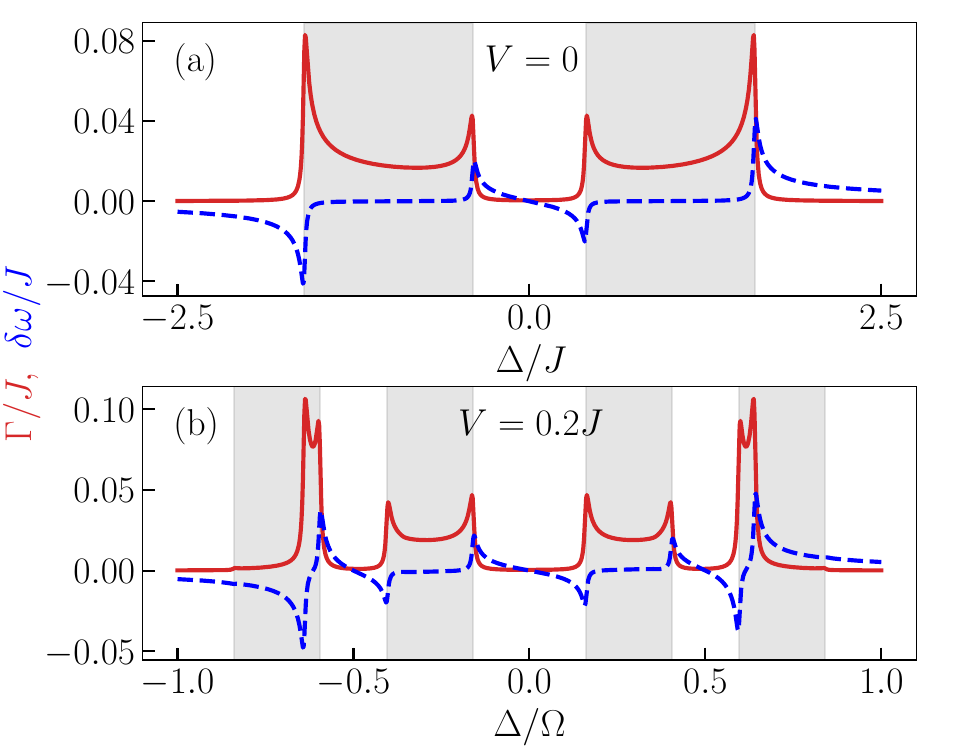}
    \caption{Decay rate and Lamb shift for an emitter coupled to the $A$-sublattice of a photonic bath with driving amplitudes (a) $V=0$ and (b) $V=0.2J$, respectively. Shaded gray depicts when the detuning $\Delta$ lies in the band regime. The other parameters are: $J'=0.6J$, $\Omega=2.5J$, $g=0.1J$.}
    \label{fig:selfenergy}
\end{figure}

In this section, we study the interactions between quantum emitters mediated by the time-dependent one-dimensional lattice presented in Section~\ref{subsec: photonic bath}. For that, we apply the same transformations and get an effective light-matter interaction Hamiltonian in the regimes of interest for this manuscript.
This effective Hamiltonian will be useful to understand, from analytical expressions, some of the features of the emergent interactions. Then, in section~\ref{subsec:single} we analyze the single emitter limit, by studying both, the modification of its dynamics and the appearance of emitter-photon bound states in the $0$- and the $\pi$-gaps. Finally, in section~\ref{subsec:many}, we consider the many emitter scenario, and demonstrate the emergence of perfect coherent exchanges between emitters coupled to the bath.

\subsection{Effective light-matter Hamiltonian~\label{subsec:effectiev}}

By applying the same transformations and approximations as in the previous section, we can write the Hamiltonian as $\hat{H}(t)\approx\hat{H}_S+\hat{H}_B^{\mathrm{RWA}}+\hat{H}_{\mathrm{int}}(t)$, where the time dependence is shifted from the bath to the interaction part of the Hamiltonian (see Appendix~\ref{App:effective Hamiltonian} to get the full details of the derivation). To see explicitly the terms that dominate this interaction, it is convenient to write $H(t)$ in the interaction picture with respect to $\hat{H}_S+\hat{H}_B^{\mathrm{RWA}}$, where it reads:
\begin{align}
    \hat{H}^I_\mathrm{int}(t) &= \frac{g}{\sqrt{2N}}\sum_{n,r,k} \Big(F^\alpha_{n,r,k}e^{ikj_n} \hat{\sigma}_n^\dagger  \hat{\mathcal{O}}_{r,k} e^{i(\omega_n-\omega_c-\lambda_r) t}  \notag\\
    &+ F_{n,r,k}^{\alpha,*} e^{-ikj_n} \hat{\sigma}_n^\dagger  \hat{\mathcal{O}}_{r,k}^\dagger e^{i(\omega_n+\omega_c+\lambda_r) t} +\mathrm{H.c.}\Big)\,,\label{eq:Htimed2}
\end{align}
and $r=1\sim 8$ is an index which indicates the $8$ possible interaction terms, and $\omega_c+\lambda_r$ and $\frac{g}{\sqrt{2N}}F^\alpha_{n,r,k} e^{ikj_n}$ are the corresponding bath operator frequencies and coupling strengths, respectively, with $\alpha$ denoting the sub-lattice that the emitter is originally coupled to. 
When $r$ is an odd number, the operator $\hat{\mathcal{O}}_{r,k}$ corresponds to $\hat{p}_k$. Instead, when $r$ is an even number, the operator  $\hat{\mathcal{O}}_{r,k}$ corresponds to $\hat{q}_k$ (details in Appendix~\ref{App:effective Hamiltonian}).

 In principle, all of the different $r$-terms can contribute, however, we will focus in a parameter regime where we can make some simplifications. First, the cavity frequency $\omega_c$ is much larger than the hoppings $J$ and $J'$, and we are interested in intermediate-frequency regime $\Omega\in 2\omega(k)\sim J$ where $\lambda_r \ll \omega_c$. Thus, the detunings appearing in Eq.~\eqref{eq:Htimed2} approximately reduce to $\pm(\omega_n-\omega_c)$ and $\pm(\omega_n+\omega_c)$. Besides, we also work in the weak coupling limit, i.e., $g\ll \omega_n,\omega_c$, which allows us to neglect the rapidly oscillating terms which do not conserve the number of excitations. The effective Hamiltonian in the interaction picture is written as
\begin{align}\label{eq:H_eff}
    \hat{H}_\mathrm{eff}(t) &= \frac{g}{\sqrt{2N}}\sum_{n,r,k} F^\alpha_{n,r,k}e^{ikj_n} \hat{\sigma}_n^\dagger  \hat{\mathcal{O}}_{r,k} e^{i(\omega_n-\omega_c-\lambda_r) t} +\mathrm{H.c.}\,.
\end{align}

In the subsequent discussion, all analytical results are based on the effective Hamiltonian in Eq.~\eqref{eq:H_eff}, whereas the exact numerical simulations will be based on the full light-matter interaction Hamiltonian with the excitation non-conserving terms dropped: 
\begin{align} \label{eq:H_num}
    \hat{H}_\mathrm{full}(t)=\hat{H}_S + \hat{H}_B(t)+\hat{H}_{\mathrm{int}}^{\mathrm{RWA}},
\end{align}
for $\hat{H}^{\mathrm{RWA}}_\mathrm{int} = g\sum_{n} \left(\hat{\sigma}_n^\dagger \hat{\alpha}_{j_n}  + \mathrm{H.c.}\right)$.

\subsection{Single emitter dynamics and bound-states~\label{subsec:single}}

To start gaining intuition, let us first study what happens when a single emitter with detuning $\Delta=\omega_0-\omega_c$ couples to the lattice at position $j_0$. In the static SSH situation, as thoroughly considered in Ref.~\cite{Bello2019a}, the dynamics of the emitters is fully determined by the self-energy defined as:
\begin{align}
    \Sigma_e(z)=\sum_{k,\alpha=u,l} \frac{|g_{\alpha,k}|^2}{z-\omega_\alpha(k)}\, ,
\end{align}
where $g_{\alpha,k}$ is the coupling strength to eigenmode $\hat{\alpha}_k=\hat{u}_k,\hat{l}_k$.  In the weak coupling (Markov) limit, the  real and imaginary parts, of the self-energy, $\Sigma_e(\Delta+i0^+)=\delta\omega(\Delta)-i\Gamma(\Delta)/2$, encode the information of the renormalization of the emitters' frequency (Lamb-shift) and lifetime. In Fig.~\ref{fig:selfenergy}(a), we plot the Lamb shift (dashed blue line) and decay rate (solid red lines) for a range of emitter's frequencies $\Delta\in\left[-5J,\, 5J\right]$ in the static situation $V=0$. There, we observe how $\Gamma(\Delta)\neq 0$ for all the frequency regions overlapping with the static bands $\omega(k)$ (shaded gray). When $\Delta\notin \pm\omega(k)$, it can be shown that $\Gamma(\Delta)\equiv 0$, because there are no modes available at those frequencies.  In those cases, the excitation of the emitter can be shown to localize around it forming what has been called as emitter-photon bound-state~\cite{bykov75a,john90a,kurizki90a}.
The energy and spatial shape of this bound state can be found by solving the time-independent Schr\"odinger equation $\hat{H}\ket{\Psi_\mathrm{BS}}=E_\mathrm{BS}\ket{\Psi_\mathrm{BS}}$, in the single excitation subspace:
\begin{align}
   \ket{\Psi_\mathrm{BS}}=\left( C_e \ket{e}\otimes+\ket{g}\otimes\sum_{j, \,\alpha=a,b}C_{\alpha,j}\hat{\alpha}^\dagger_j\right)\ket{\mathrm{vac}}_B\,,
\end{align}
which in the static case can be done analytically finding~\cite{Bello2019a}:
\begin{align}
   C^A_{a,j}=\frac{g C_e}{2\pi} \int_{-\pi}^\pi dk\, \frac{E_{\text{BS}}e^{ikj}}{E_{\text{BS}}^2-\omega^2(k)}\,,\label{eq:BS1}\\
    C^A_{b,j}=-\frac{g C_e}{2\pi} \int_{-\pi}^\pi dk\, \frac{\omega(k)e^{i(kj-\theta_k)}}{E_{\text{BS}}^2-\omega^2(k)}\,,\label{eq:BS2}
\end{align}
where the superscript $A$ represents that the emitter couples to sublattice $A$. These bound states are exponentially localized to the left/right and feature support in the $B/A$ sublattice depending on whether they couple to the $A/B$ sublattices, respectively~\cite{Bello2019a,Leonforte2020b}. The origin of this behaviour can be traced back to the topological edge states of the two semi-infinite chains "broken" by the emitters's coupling.

\begin{figure}[tb]
    \centering
   \includegraphics[width=0.95\linewidth]{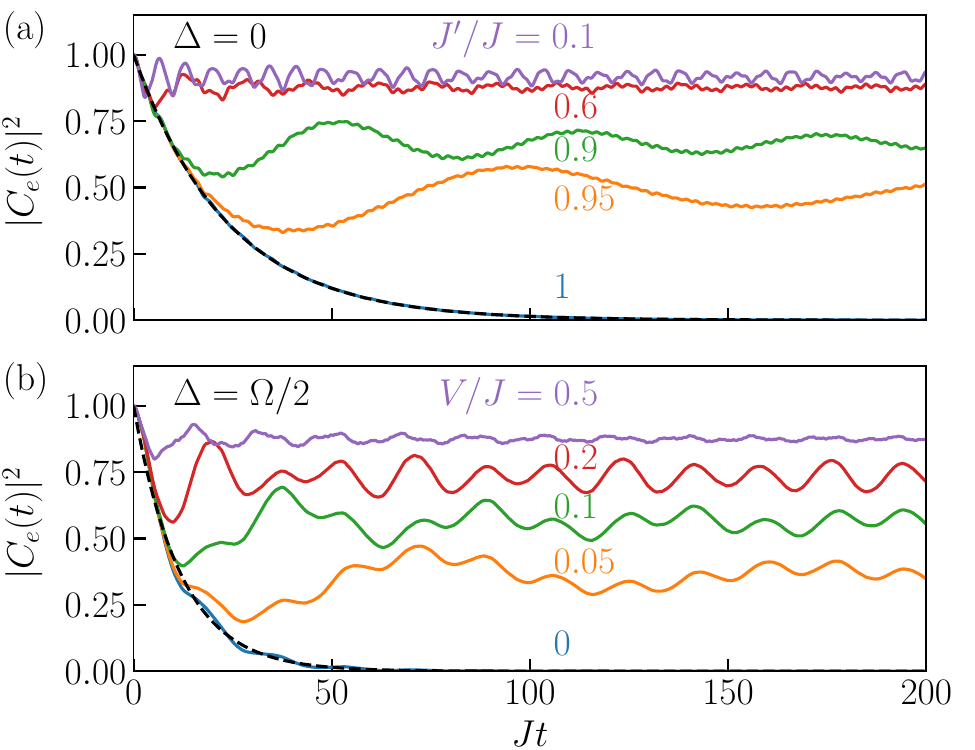}
    \caption{Exact population dynamics $|C_e(t)|^2$ for a single emitter coupled to the A sub-lattice of a chain with PBC and $N=200$ cells. (a) In the $0$-gap $\Delta=0$ with $V=0.2J$ and (b) In the $\pi$-gap $\Delta=\Omega/2$ with $J'=0.6J$. In colors we plot the exact dynamics for different parameters, while in dashed black we plot the exponential decay predicted by the Markov approximation of Eq.~\eqref{eq:self_energy} in the cases where the gap closes (solid blue).  Other parameters: $\Omega=2.5J$, $g=0.2J$.}
    \label{fig:dynamics}
\end{figure}

Now, let us see what happens with these predictions when we consider a time-dependent situation. For that, we plot in  Figs.~\ref{fig:dynamics}(a) and (b) the decay of an initially excited emitter, $\ket{\Psi(0)}=\ket{e}\otimes\ket{\mathrm{vac}}_B$, when it couples to a time-dependent bath and its splitting is tuned to the $0$- and $\pi$-gaps, respectively. In the case of weak field $V\ll \Omega $, the opening and closing of the $0$-gap is controlled by $J'/J$, as in the static case~\cite{Bello2019a}. The larger $J'/J\gg 1$, the larger is the $0$-gap and the dynamics is closer to the Markovian prediction where the emitter is not decaying. As $J'/J\rightarrow 1$, the band-gap gets smaller and non-Markovian effects become more prominent. In particular, in Fig.~\ref{fig:dynamics}(a), we see that as $J'/J$ get closer to $1$, the fractional decay behavior becomes more noticeable.
In the $J'=J$ case (in solid blue), the decay becomes purely exponential because the band-gap closes, matching exactly the Markovian prediction depicted in dashed black.  In Fig.~\ref{fig:dynamics}(b), we plot the situation with fixed $J'=2J$ and $\Delta=\Omega/2$ for different driving amplitudes $V$. In that case, if $V=0$ (solid blue), we see that the emitter features an exponential decay because at that frequency, the emitter is resonant to one of the static bands, see Fig.~\ref{fig:model}(b). However, as soon as $V\neq 0$, the $\pi$-gap opens and the system starts featuring fractional decay dynamics. 

These numerical observations from Fig.~\ref{fig:dynamics} can be explained by calculating the renormalized self-energy operator using the effective light-matter couplings of Eq.~\eqref{eq:H_eff} as follows:
\begin{align}
    \Sigma_{\mathrm{eff}}(z) = \frac{g^2}{2N}\sum_{r,k} \frac{|F^\alpha_{n,r,k}|^2}{z-\lambda_r}\,.\label{eq:self_energy}
\end{align}

We note that in this expression we have used the RWA to ignore the terms that oscillate with the field frequency $\Omega$, and thus obtain time-independent self-energy. In Fig.~\ref{fig:selfenergy}(b), we plot the real and imaginary part of this expression as a function of the emitter's detuning, $z=\Delta+i 0^+$, where we see clearly the appearance of new regions where $\Gamma(\Delta)\equiv 0$ due to the drive-induced $\pi$-gaps. 
Besides, we observe that non-zero decay occurs predominantly at the static band frequency rather than across all Floquet bands. This is because, in the weak driving regime, the different sidebands is minimal, resulting in nearly all of the spectral weight being concentrated near the static band.

\begin{figure}[tb]
    \centering   \includegraphics[width=0.95\linewidth]{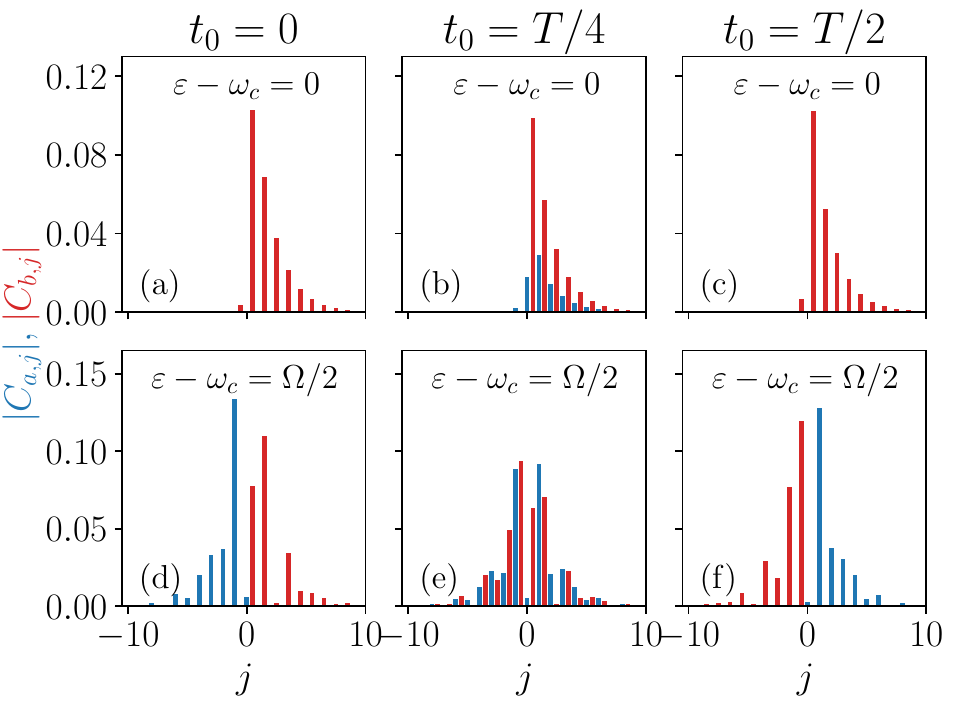}
    \caption{Spatio-temporal shape of the bound states for $\Delta=0$ (top row) and $\Delta=\Omega/2$ (bottom row). Probability amplitudes $|C_{a,j}|$ are shown in blue, while the $|C_{b,j}|$ are shown in red. Parameters: $J'=0.6J$, $V=0.2J$, $\Omega=2.5J$, $g=0.1J$.}
    \label{fig:bound_states}
\end{figure}

Finally, let us mention that the fractional decay observed in Fig.~\ref{fig:dynamics}(b) points to the emergence of emitter-photon bound states associated to the emergence of these new band-gaps. We confirm this intuition by analyzing the quasi-energy spectrum of the Hamiltonian of Eq.~\eqref{eq:H_num} (not shown), where we find emitter-photon bound states in both the $0$- and $\pi$-gaps. In Fig.~\ref{fig:bound_states}, we plot the corresponding spatial shape of the bound-states of the $0$- (upper row) and $\pi$-gaps (lower row) at different times (in the different columns). There, we see how the bound state of the $0$-gap (upper row) has very similar properties to the one found in the static SSH model: it is localized preferentially to the right and only in the $B$ sublattice. In intermediate times, there is a small population transfer to the $A$ sublattice, but this will be reduced as the size of the $\pi$-gap decreases.
On the contrary, the bound-state appearing when the emitter is resonant with the $\pi$-gap, i.e., $\Delta=\Omega/2$, has several differences with the static case:
\begin{itemize}
    \item It has a non-monotonic spatial dependence, like the anomalous edge states of the $\pi$-gap, shown in Fig.~\ref{fig:edge_state}.
    \item It changes with time. For example, at $t_0=0$, the localization at both sides of the emitters occurs in different sublattices; at $t_0=T/4$, it localizes in both sublattices, whereas at $t_0=T/2$ it reverses the original shape.
\end{itemize}

In the next section, we will see how such time-dependent bound-states are able to mediate interactions between emitters like in the static case. 


\subsection{Coherent excitation exchanges between emitters~\label{subsec:many}}
\begin{figure}[tb]
    \centering
   \includegraphics[width=1\linewidth]{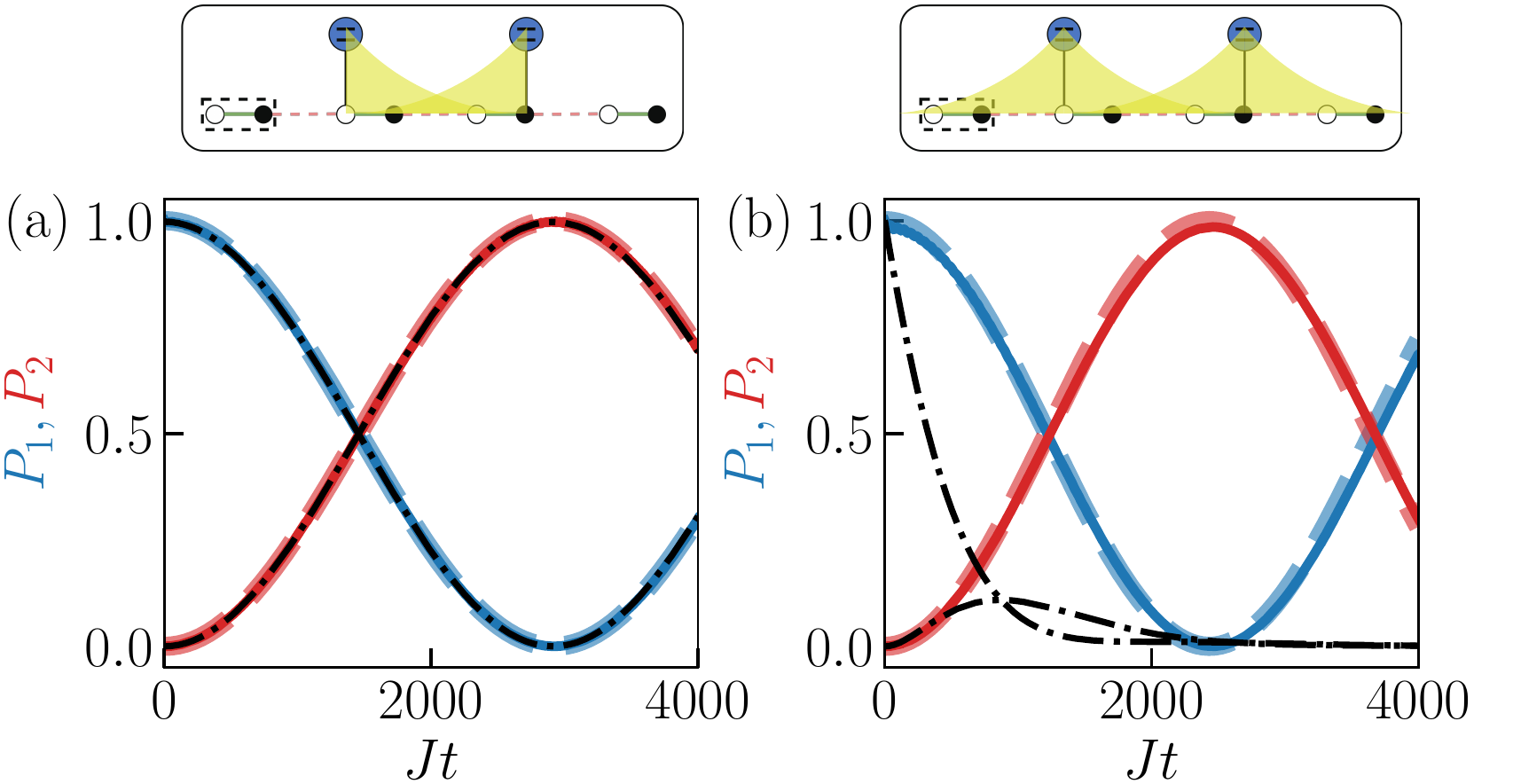}

    \caption{Coherent exchanges for emitter $1$ in sublattice $A$ and emitter $2$ in sublattice $B$ with distance $j_{12}=-1$. (a) $\Delta=0$. (b) $\Delta=\Omega/2$. Solid lines represent the exact solution and wide dashed lines for analytical solution, and black dotted-dashed lines for static case. Other parameters: $J^\prime=0.6J$, $V=0.1J$, $\Omega=2.5J$, $g=0.03J$.}
    \label{fig:exchange}
\end{figure}

Let us consider now that we have two emitters with detunings $\Delta_n=\omega_n-\omega_c$ coupled to the time-dependent bath. In the static case, it is well-known that the existence of emitter-photon bound states results in a perfect coherent excitation exchange between two emitters with same detuning $\Delta_1=\Delta_2=\Delta$. This can be captured in the Markov limit by a dipole-dipole interaction Hamiltonian:
\begin{align}
\hat{H}_\mathrm{QE}=\sum_{n,m}G_{nm}^{\alpha\beta}\hat{\sigma}_n^\dagger \hat{\sigma}_m\, ,\label{eq:Hdipdip}
\end{align}
where $G_{nm}^{\alpha\beta}$ is given by the overlap between the emitter-photon bound states with the different emitters given by Eqs.~\eqref{eq:BS1}-\eqref{eq:BS2}. For this reason, when $\Delta=0$, two emitters within the same sub-lattice will never couple since the emitter-photon bound state is at $E_\mathrm{BS}$ and thus only have weights in the opposite sublattice. However, when the emitters couple to different sublattices, they exchange interactions with a frequency rate $(G_{nm}^{\alpha\beta})^{-1}$.

Let us now see what happens in the time-dependent bath situation by investigating the dynamics of two emitters when one of them is initially excited i.e., $\ket{\Psi(0)}=\ket{e}_1\otimes\ket{g}_2\otimes\ket{\mathrm{vac}}_B$. This is what we show in Fig.~\ref{fig:exchange}, where we plot the population of the initially excited emitter (in solid blue), and the other one (in solid red) as a function of time, calculated exactly solving the time-dependent Schr\"odinger equation for the Hamiltonian in Eq.~\eqref{eq:H_num}. The two panels (a) and (b) correspond to the situations where the emitter lies in the $0$- and $\pi$-gaps, respectively. In Fig.~\ref{fig:exchange}(a), we see how the bound state mediates a perfect excitation exchange. There, we also plot the results corresponding to a static case situation, i.e., with $V=0$, in dotted-dashed black, showing how the dynamics of the $0$-gap is indeed dominated by the bound state physics of the static situation.

This is very different from what happens in Fig.~\ref{fig:exchange}(b) when the emitters' frequencies lie in the $\pi$-gap, i.e., $\Delta=\Omega/2$. In that case, the exact dynamics (in solid colors) also leads to perfect coherent exchanges, while the dynamics of the equivalent static situation, i.e., fixing $V=0$, (dashed black) leads to dissipative dynamics and no oscillations. The reason is that there is no band-gap at these frequencies for the static case, which shows the temporally-induced origin of these oscillations. Interestingly, we can use our effective Hamiltonian of Eq.~\eqref{eq:H_eff} to derive an effective dipole-dipole interaction Hamiltonian like the one of Eq.~\eqref{eq:Hdipdip} for this time-dependent situation, as shown in detail in Appendix~\ref{app:meq}. From this derivation, one can obtain an effective time-independent exchange rate given by:
\begin{align}
G^{\alpha\beta}_{nm}= \frac{g^2}{2N}\sum_{r,k} \frac{F^\alpha_{n,r,k} F^{\beta,*}_{m,r,k}}{\Delta-\lambda_r} e^{ikj_{nm}}\,,\label{eq:dipdip2}
\end{align}
where $j_{nm}=j_n-j_m$ is the distance between $n$-th emitter and $m$-th emitter. The dynamics obtained through this effective master equation is what we plot in dashed colors in Fig.~\ref{fig:exchange}, showing indeed a perfect quantitative agreement with the dynamics obtained with the exact calculation

\begin{figure}[tb]
    \centering
   \includegraphics[width=0.95\linewidth]{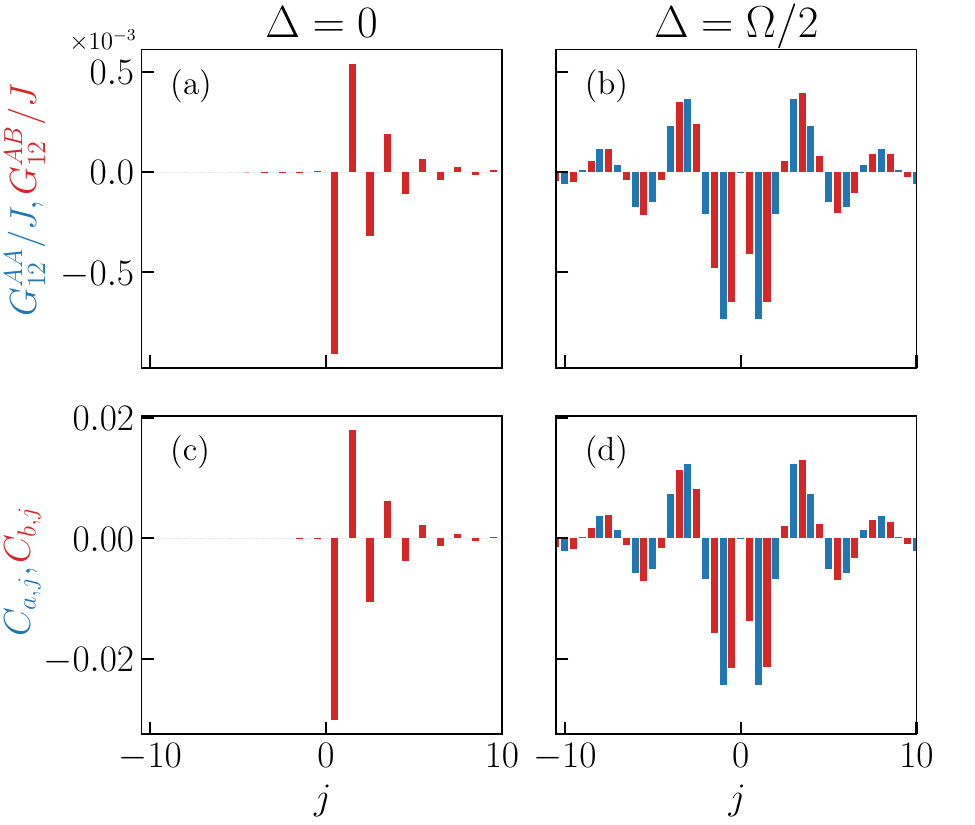}

     \caption{Comparison between analytical solution of the dipole-dipole interactions $G_{12}^{AA}$ and $G_{12}^{AB}$ (the top row, the emitter 1 couples to sublattice $A$ in $j=0$ cell.) obtained from Eq.~\eqref{eq:dipdip2} with the photon distribution of time-averaged bound state $\overline{\ket{\Phi_\mathrm{BS}}}$ (the bottom row). Parameters: $J^\prime=0.6J$, $V=0.1J$, $\Omega=2.5J$, $g=0.03J$.}
    \label{fig:coupling}
\end{figure}

To gain more intuition of these coherent exchanges, we plot their spatial distribution in Fig.~\ref{fig:coupling}(a-b) for the $0$- and $\pi$-gaps, respectively. There, we see very clearly that the shape of the $0$-gap situation has indeed the same qualitative shape than the static situation, as expected from all our previous results. On the contrary, panel (b) corresponding to the $\pi$-gap shows a very different behavior, featuring a non-monotonic spatial decay. 
It is also apparently very different from the bound state shapes found for a single emitter, which we show in the lower row of Fig.~\ref{fig:bound_states}. However, if we time-average the bound state in those figures, i.e., $\overline{\ket{\Phi_{\mathrm{BS}}}}=\int_0^T dt \,\ket{\Phi_{\mathrm{BS}}(t)} /T$, and plot their corresponding averaged spatial distribution, as we do in Figs.~\ref{fig:coupling}(c-d), we indeed find that they feature the same shape as $G_{nm}^{\alpha\beta}$. This explains that in the case of a time-dependent bath, the interaction is mediated by a time-averaged version of the bound-states, rather than by its full time-dependent version.

Let us finally also show the potential of time-dependent baths to obtain coherent exchanges in situations that would be impossible for the static baths. In particular, it is well known that if the emitters have different frequencies in the static case, $\Delta_1\neq \Delta_2$, the coherent exchange is precluded, completely disappearing when $|\Delta_1-\Delta_2|\gg G_{12}^{\alpha\beta}$. However, let us now show that, in the time-dependent situation, this energy difference can be compensated by the driving field. 

\begin{figure}[tb]
    \centering   \includegraphics[width=1\linewidth]{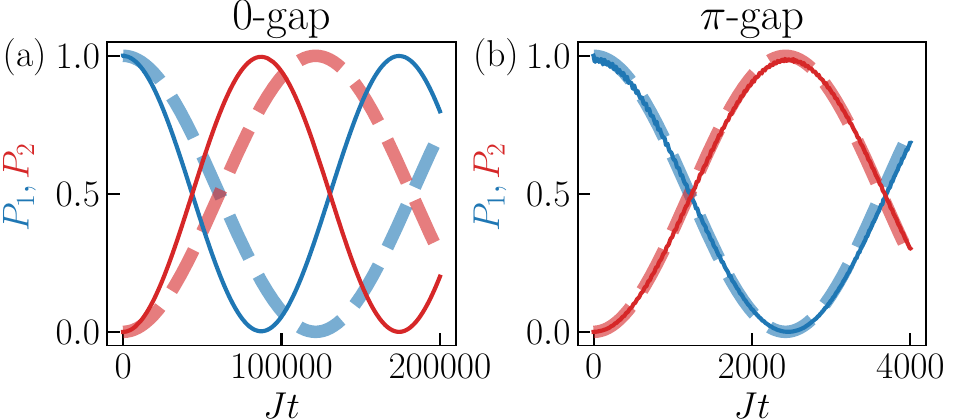}
    \caption{Coherent exchanges for emitter $1$ in sublattice $A$ and emitter $2$ in sublattice $B$ with distance $j_{12}=-1$.
    (a) $\Delta_1=\Omega -\delta \omega(\Omega),\Delta_2=0$ and (b) $\Delta_1=-\Delta_2=\Omega/2$.
    Solid lines represent the exact solution and wide dashed lines for analytical solution of Eq.~\eqref{eq:effOm}.    Other parameters: $J^\prime=0.6J$, $V=0.1J$, $\Omega=2.5J$, $g=0.03J$.}
    \label{fig:exchange_Omega}
\end{figure}

To show that, in Fig.~\ref{fig:exchange_Omega}(a-b) we plot the excitation exchanges for two emitters with different frequencies in both the $0$- and $\pi$-gaps, respectively, calculated using the exact dynamics. In both cases, one can see how the drive can indeed compensate the energy difference, and obtain perfect coherent exchanges. For the $0$-gap, we need to have a small correction to the bare emitter's energies, i.e., $\Delta_1=\Omega-\delta\omega(\Omega)$ and $\Delta_2=0$, with $\delta\omega(\Omega)=\mathrm{Re}[\Sigma_\mathrm{eff}(\Omega+i0^+)]$. For the $\pi$-gap, the perfert coherent exchanges occur directly for $\Delta_1=-\Delta_2=\Omega/2$. To understand better why we need to introduce this correction compared to the equal frequencies situation, we derive in Appendix~\ref{app:meq} an effective Hamiltonian that describes the excitation exchanges in this regime. In the interaction picture, this effective Hamiltonian reads:
\begin{align}
    \hat{H}_\mathrm{QE} \approx &\, \sum_{n}\delta \omega_n \hat{\sigma}_n^\dagger \hat{\sigma}_n + G^{\alpha\beta,\Omega}_{12}\left(\hat{\sigma}_1^\dagger  \hat{\sigma}_2e^{i(\Delta_1-\Delta_2-\Omega)t} + \mathrm{H.c.}\right)\,,\label{eq:effOm}
\end{align}
where we consider the possibility of different Lamb-shift for the two emitters due to their different initial energies, i.e., $\delta\omega_n=\delta\omega(\Delta_n)=\mathrm{Re}[\Sigma_\mathrm{eff}(\Delta_n+i0^+)]$. In the $\pi$-gap, both Lamb-shifts are the same, i.e., $\delta\omega(\pm\Omega/2)=0$, and thus there is no influence on the transfer. However, in the $0$-gap, $\Delta_2=0$ is not shifted because $\delta\omega(\Delta_2)=0$, whereas $\delta\omega(\Delta_1)\neq 0$. Thus, it needs to be compensated to observe a perfect population transfer, like we do in Fig.~\ref{fig:exchange_Omega}(a). From this derivation, we obtain a semi-analytical expression for the effective coupling strength induced by the bath that reads:
\begin{align} 
G_{12}^{\alpha\beta,\Omega}&=\frac{g^2}{2N}\sum_{k,r=1}^{r=6}    \frac{F^{\alpha}_{1,r,k} F_{2,r+2,k}^{\beta,*}}{\Delta_2-\lambda_{r+2}} e^{ikj_{12}}\,.
\end{align}

When we use this expression in the $\pi$-gap, the dynamics given by the effective Hamiltonian of Eq.~\eqref{eq:effOm} exactly matches the exact solution, as we show in Fig.~\ref{fig:exchange_Omega}(b), in dashed lines. 
In the $0$-gap, there is a frequency mismatch because the terms neglected to arrive to the RWA expression in Eq.~\eqref{eq:effOm} are not sufficiently small. This produces an effective frequency shift, which leads to a deviation  between the analytical and numerical solution which is more evident as time increases (more details in Appendix~\ref{app:meq}.

\section{Conclusion~\label{sec:conclusion}}

In summary, we have made a first systematic study of the interactions between quantum emitters mediated by a time-dependent one-dimensional structured photonic lattice described by an AC-driven Su-Schrieffer-Heeger (SSH) model. By leveraging Floquet theory, we have demonstrated the emergence of topological bound states in the quasi-energy spectrum, which mediate robust, tunable interactions between emitters. These bound states, appearing in both the 0-gap and \(\pi\)-gap, offer unique features that have no static counterpart, including their ability to induce coherent exchanges in frequency regions that would otherwise be highly dissipative in static systems or where the exchanges would not occur due to frequency emitter's mismatch.

Our analysis highlights the versatility of time-dependent photonic media in engineering non-trivial photon-mediated interactions, extending beyond the capabilities of static structured baths. These results pave the way for exploring higher-dimensional Floquet photonic lattices and more complex emitter configurations, opening prospects for advanced quantum simulation, metrology, and information processing. Future work could focus on experimental implementations of these systems and their extension to high-dimensional systems.

\begin{acknowledgments}
The authors acknowledge support from the CSIC Research Platform on Quantum Technologies PTI-001 and from Spanish projects PID2021-127968NB-I00 and PID2023-146531NA-I00, funded by MICIU/AEI/10.13039/501100011033/ and by FEDER Una manera de hacer Europa, TED2021-130552B-C22 funded by  MICIU/AEI /10.13039/501100011033 and by the European Union NextGenerationEU/ PRTR, respectively, and from the QUANTERA project MOLAR with reference PCI2024-153449 and funded MICIU/AEI/10.13039/501100011033 and by the European Union, and supported by the National Natural Science Foundation of China under Grants No. W2411002 and No. 12375018.
\end{acknowledgments}

\appendix

\begin{widetext}

\section{Derivation of the effective time-independent, light-matter Hamiltonian~\label{App:effective Hamiltonian}}

The full light-matter interaction Hamiltonian is composed by three terms: $\hat{H}(t)=\hat{H}_B(t)+\hat{H}_S+\hat{H}_\mathrm{int}$. For the bath, we consider a photonic lattice described by an AC-driven Su-Schrieffer-Heeger (SSH). The Hamiltonian of this bath is then composed by a time-independent and a time-dependent part, i.e., $\hat{H}_B(t) = \hat{H}_\mathrm{SSH} + \hat{H}_d(t)$, where:
\begin{align}
\hat{H}_\mathrm{SSH}&= \sum_j\left[\omega_c \hat{a}^\dagger_j \hat{a}_j + \omega_c \hat{b}^\dagger_j \hat{b}_j +  \left( J \hat{a}_j^\dagger \hat{b}_j +   J' \hat{a}_{j+1}^\dagger \hat{b}_j +  \mathrm{H.c.} \right)\right]\,,  \\ \notag
    \hat{H}_d(t)&= \sum_j \left[2V\cos(\Omega t)\left(  \hat{a}_j^\dagger \hat{b}_j - \hat{a}_{j+1}^\dagger \hat{b}_j\right) +  \mathrm{H.c.}\right] \,,
\end{align}

Thus, $J_1(t)=J+2V\cos(\Omega t)$ and $J_2(t)=J'-2V\cos(\Omega t)$ are the full time-dependent, intra-cell and inter-cell hoppings, respectively.

The emitters and light-matter interaction Hamiltonians are given by:
\begin{align}
    \hat{H}_S &=\sum_n\omega_n \hat{\sigma}_n^\dagger \hat{\sigma}_n\,, \\ \notag
    \hat{H}_\mathrm{int}&= g\sum_{n\in S_A}\left(\hat{\sigma}_n+\hat{\sigma}_n^\dagger\right) \left(\hat{a}_{j_n} + \hat{a}_{j_n}^\dagger \right) + g\sum_{n\in S_B} (\hat{\sigma}_n+ \hat{\sigma}_n^\dagger) \left(\hat{b}_{j_n} + \hat{b}_{j_n}^\dagger  \right)\,.
\end{align}

Assuming periodic boundary conditions (PBC), one can define the following Fourier transform of the bath operators $\hat{a}_k [\hat{b}_k] = \frac{1}{\sqrt{N}} \sum_j \hat{a}_j [\hat{b}_j] e^{-i k j}$, which we can rewrite the Hamiltonians in momentum space:
\begin{align}
    \hat{H}_\mathrm{SSH} &=  \sum_k \left[\omega_c \hat{a}^\dagger_k \hat{a}_k + \omega_c \hat{b}^\dagger_k \hat{b}_k +\left(h_k\hat{a}_k^\dagger \hat{b}_k +\mathrm{H.c.}\right)\right]\,, \\ \notag
    \hat{H}_d(t)&=  \sum_k \left[2d_k\cos(\Omega t)\hat{a}_k^\dagger \hat{b}_k +\mathrm{H.c.}\right]\,, \\ \notag
    \hat{H}_\mathrm{int}&= \frac{g}{\sqrt{N}} \sum_{k,n\in S_A} \left(\hat{\sigma}_n+\hat{\sigma}_n^\dagger\right) \left(e^{ikj_n} \hat{a}_k + \mathrm{H.c.}\right) + \frac{g}{\sqrt{N}} \sum_{k,n\in S_B} \left(\hat{\sigma}_n+\hat{\sigma}_n^\dagger\right) \left(e^{ikj_n} \hat{b}_k + \mathrm{H.c.}\right)\,,
\end{align}
where $h_k=J+J'e^{-ik}=\omega(k) e^{i\theta_k}$ and $d_k=V(1-e^{-ik})=|d_k| e^{i\beta_k}$ are time-independent and time-dependent hopping amplitude, respectively.

Now, we can define the following unitary transformation to cancel the time-dependent part $\hat{H}_d(t)$:
\begin{align}
&\hat{R}_d(t)=e^{-i\int \hat{H}_d(t)\,dt}=e^{-i\sum_k\left[ \frac{2d_k}{\Omega}\sin(\Omega t) \hat{a}_k^\dagger \hat{b}_k +  \mathrm{H.c.}\right]}\,.
\end{align}

Defining $\mathbf{V}_k^\dagger = (\hat{a}_k^\dagger,\, \hat{b}_k^\dagger)$, we can rewrite the unitary transformation as:
\begin{align}
    &\hat{R}_d(t)=\sum_k \mathbf{V}_k^\dagger R_d(k,t)\mathbf{V}_k\,, \\ \notag
    &R_d(k,t)=e^{-if_k(t)  M_k} \,,\\ \notag
    &M_k = 
    \left(
    \begin{array}{cc}
    0 & e^{i\beta_k}  \\ 
    e^{-i\beta_k} & 0
    \end{array}
    \right)
    = \cos(\beta_k)\sigma_x -\sin(\beta_k) \sigma_y\,,
\end{align}
where $f_k(t)=2|d_k|\sin(\Omega t)/\Omega=z_k\sin(\Omega t)$.
$\sigma_x$ and $\sigma_y$ are the Pauli matrices. Expanding $R_d(k,t)$ in matricial form, we obtain:
\begin{align}
    R_d(k,t)&=e^{-if_k(t) ( \cos(\beta_k)\sigma_x -\sin(\beta_k) \sigma_y)}=\cos\left(f_k(t)\right) \mathcal{I} -i \sin\left(f_k(t)\right) ( \cos(\beta_k) \sigma_x -\sin(\beta_k) \sigma_y) \\ \notag
    &=\left(
    \begin{array}{cc}
    \cos\left(f_k(t)\right)& -i\sin\left(f_k(t)\right)e^{i\beta_k} \\ 
    -i\sin\left(f_k(t)\right)e^{-i\beta_k} &  \cos\left(f_k(t)\right)
    \end{array}
    \right)\,,
\end{align}
where $\mathcal{I}$ is the unit matrix. Now, we can obtain the operators under this time-dependent unitary transformation as follows, $\hat{\mathcal{O}}_k(t)=\hat{R}^\dagger_d(t) \hat{\mathcal{O}}_k\hat{R}_d(t)$. Doing that for the bath operators, we find:
\begin{align}
    \left(
    \begin{array}{cc}
    \hat{a}_k(t) \\
    \hat{b}_k(t)
    \end{array}
    \right)
    = R_d^\dagger (k,t) \mathbf{V_k} =
    \left(
    \begin{array}{cc}
    \cos(f_k(t)) \hat{a}_k + i\sin(f_k(t))e^{i\beta_k} \hat{b}_k  \\ 
     \cos(f_k(t))  \hat{b}_k + i\sin(f_k(t))e^{-i\beta_k} \hat{a}_k
    \end{array}
    \right)\,.
\end{align}

Substituting this expression into the bath Hamiltonian, we get:
\begin{align}
    \hat{H}^d(t) &= \hat{R}_d^\dagger(t) \hat{H}(t)\hat{R}(t)-i\hat{R}_d^\dagger(t) \dot{\hat{R}}_d(t)=\hat{H}_B^d(t)+\hat{H}_S+\hat{H}_\mathrm{int}^d(t)\,,
\end{align}
where
\begin{align}
    \hat{H}_B^d(t) &=\hat{R}_d^\dagger (t) \hat{H}_\mathrm{SSH} \hat{R}_d(t)\\ \notag
    &=\sum_k \left[\left\{\omega_c +  \mathrm{Re}[ie^{-i\beta_k}h_k] \sin(2f_k(t))\right\}\hat{a}_k^\dagger \hat{a}_k  +  \left\{\omega_c - \mathrm{Re}[ie^{-i\beta_k}h_k] \sin(2f_k(t))\right\}  \hat{b}_k^\dagger \hat{b}_k \right] \\ \notag
    &+ \sum_k\left[ \frac{1}{2}\left\{h_k[1+\cos(2f_k(t))]\ +  e^{2i\beta_k}h_k^* [1-\cos(2f_k(t)) ] \right\} \hat{a}_k^\dagger \hat{b}_k + \mathrm{H.c.}\right]\,,
\end{align}
\begin{align}
    \hat{H}_\mathrm{int}^d(t)&= \hat{R}_d^\dagger (t) \hat{H}_\mathrm{int} \hat{R}_d(t)\\ \notag
    &= \frac{g}{\sqrt{N}} \sum_{k,n\in S_A} \left[e^{ikj_n}(\hat{\sigma}_n^\dagger+\hat{\sigma}_n)\left(\cos(f_k(t)) \hat{a}_k + i\sin(f_k(t))e^{i\beta_k}\hat{b}_k\right) + \mathrm{H.c.}\right]\\ \notag
    &+\frac{g}{\sqrt{N}} \sum_{k,n\in S_B}\left[ e^{ikj_n}(\hat{\sigma}_n^\dagger+\hat{\sigma}_n)\left(\cos(f_k(t))\hat{b}_k + i\sin(f_k(t))e^{-i\beta_k}\hat{a}_k\right) + \mathrm{H.c.}\right]\,.\\ \notag
\end{align}

To simplify further the expressions, we can use the Jacobi-Anger expansion of the trigonometric functions, which read:
\begin{align}
    \cos\left(z\sin(\theta)\right) 
    &= \mathcal{J}_0(z) + 2\sum_{m=1}^\infty \mathcal{J}_{2m}(z) \cos(2m\theta)\,, \\ \notag
    \sin\left(z\sin(\theta)\right) 
    &=  2\sum_{m=1}^\infty \mathcal{J}_{2m-1}(z) \sin((2m-1)\theta)\,,
\end{align}
where $\mathcal{J}_m(z)$ is the $m$-th Bessel function of the first kind. Inserting these expressions into the previous equation, we arrive to a complete expression of the bath Hamiltonian which takes into account all the possible multi-photon processes:
\begin{align}
    &\hat{H}_B^d(t) =\sum_k \left\{\omega_c +  2\mathrm{Re}[ie^{-i\beta_k}h_k] \sum_{m=1}^\infty \mathcal{J}_{2m-1}(2z_k) \sin((2m-1)\theta)\right\}\hat{a}_k^\dagger \hat{a}_k \\ \notag
    &+\sum_k  \left\{\omega_c - 2\mathrm{Re}[ie^{-i\beta_k}h_k] \sum_{m=1}^\infty \mathcal{J}_{2m-1}(2z_k) \sin((2m-1)\theta)\right\}  \hat{b}_k^\dagger \hat{b}_k  \\ \notag
    &+ \sum_k\left\{ \frac{1}{2}\left[h_k\left(1+\mathcal{J}_0(2z_k) + 2\sum_{m=1}^\infty \mathcal{J}_{2m}(2z_k) \cos(2m\theta)\right) +  e^{2i\beta_k}h_k^* \left(1-\mathcal{J}_0(2z_k) + 2\sum_{m=1}^\infty \mathcal{J}_{2m}(2z_k) \cos(2m\theta) \right) \right] \hat{a}_k^\dagger \hat{b}_k + \mathrm{H.c.}\right\}\,,
\end{align}

Since we are only interested in the intermediate-frequency regime $\Omega/2\in\omega(k)$, we will only keep terms with $\mathcal{J}_0(2z_k)$ and $\mathcal{J}_1(2z_k)$, obtaining the following effective bath Hamiltonian:
\begin{align}
    \hat{H}_B^d(t) &\approx\sum_k \left[\omega_c\hat{a}_k^\dagger \hat{a}_k + \omega_c \hat{b}_k^\dagger \hat{b}_k + \left( h_k^d\hat{a}^\dagger_k \hat{b}_k + \mathrm{H.c.}\right)\right] + \sum_k 2\gamma_k \sin(\Omega t) (\hat{a}_k^\dagger \hat{a}_k-\hat{b}_k^\dagger \hat{b}_k)\,,
\end{align}
where $h_k^d=[h_k(1+\mathcal{J}_0(2z_k))+e^{2i\beta_k}h_k^*(1-\mathcal{J}_0(2z_k))]/2=\omega_d(k)e^{i\theta_k^d}$ and $\gamma_k=\mathrm{Re}[ie^{-i\beta_k} h_k]\mathcal{J}_1(2z_k)=\mathcal{J}_1(2z_k) (J+J')V\sin(k)/|d_k|$. The first terms are time-independent, and thus can be diagonalized by the transformation $\hat{\Lambda}_1$: $\hat{u}_{k} [\hat{l}_{k}] = \frac{1}{\sqrt{2}} \left( \hat{a}_k +[-] e^{i\theta_k^d} \hat{b}_k \right)$. Using these operators, the bath Hamiltonian then reads:
\begin{align}
    \hat{H}_B^d(t)& = \sum_k \left[(\omega_c+\omega_d(k)) \hat{u}_k^\dagger \hat{u}_k + (\omega_c-\omega_d(k)) \hat{l}_k^\dagger \hat{l}_k + 2\gamma_k \sin(\Omega t)(\hat{u}_k^\dagger \hat{l}_k + \hat{l}_k^\dagger\hat{u}_k )\right]\,,\label{eq:Time-depBath1}
\end{align}
where $\omega_d(k)=|h_k^d|$. 

Now, we can go to the interaction picture with the static part of the bath Hamiltonian in Eq.~\eqref{eq:Time-depBath1}, and find the Hamiltonian:
\begin{align}
    \hat{H}_B^I(t)=\sum_k \left[\left(i\gamma_k  \hat{u}_k^\dagger \hat{l}_k e^{i(2\omega_d(k)-\Omega)t} + \mathrm{H.c.}\right) + \left(-i\gamma_k  \hat{u}_k^\dagger \hat{l}_k e^{i(2\omega_d(k)+\Omega)t} + \mathrm{H.c.}\right)\right] \, .
\end{align}

Here, we can apply a rotating-wave approximation (RWA) to eliminate the faster rotating terms with $e^{\pm i(\Omega+2\omega_d(k))t}$. Going back to the Schr\"odinger picture, and moving to a frame rotating with $\Omega$, i.e., $\hat{R}_\Omega(t) = \exp\left(-i \frac{\Omega}{2} t \sum_k \left( \hat{u}_k^\dagger \hat{u}_k - \hat{l}_k^\dagger \hat{l}_k \right) \right)$, we arrive to an effective time-independent Hamiltonian of the form:
\begin{align} \label{eq:App_RWA}
    \hat{H}_B^\mathrm{RWA}&=\sum_k \left[(\omega_c+\omega_d(k)-\Omega/2) \hat{u}_k^\dagger \hat{u}_k + (\omega_c-\omega_d(k)+\Omega/2) \hat{l}_k^\dagger \hat{l}_k + i\gamma_k \hat{u}_k^\dagger \hat{l}_k- i\gamma_k\hat{u}_k \hat{l}_k^\dagger\right]\,.
\end{align}

This time-independent Hamiltonian can now be diagonalized by a unitary transformation $\hat{\Lambda}_2$, using the following new operators $\hat{p}_k=\cos(\frac{\phi_k}{2}) \hat{u}_k + i\sin(\frac{\phi_k}{2})\hat{l}_k$ and $\hat{q}_k=\sin(\frac{\phi_k}{2}) \hat{u}_k - i \cos(\frac{\phi_k}{2})\hat{l}_k$ with $\phi_k=\text{atan}[\gamma_k, \, \omega_d(k)-\Omega/2]$ and $\lambda(k)=\sqrt{(\omega_d(k)-\Omega/2)^2+\gamma_k^2}$. With these definitions, the effective time-independent Hamiltonian now reads:
\begin{align}
    \hat{H}_B^{\mathrm{RWA}}&=\sum_k \left[ (\omega_c+\lambda(k)) \hat{p}_k^\dagger \hat{p}_k + (\omega_c-\lambda(k))  \hat{q}_k^\dagger \hat{q}_k\right]\,.\label{eqap:effH}
\end{align}

The same transformations and approximations have been applied to the interacting part of the Hamiltonian, $\hat{H}_\mathrm{int}$. In the interaction picture, this light-matter interaction Hamiltonian reads:
\begin{align}
    \hat{H}^I_\mathrm{int}(t) &= \frac{g}{\sqrt{2N}}\sum_{n,r,k}\left[F^\alpha_{n,r,k}e^{ikj_n} \hat{\sigma}_n^\dagger  \hat{\mathcal{O}}_{r,k} e^{i(\omega_n-\omega_c-\lambda_r) t}+ F_{n,r,k}^{\alpha,*} e^{-ikj_n} \hat{\sigma}_n^\dagger  \hat{\mathcal{O}}_{r,k}^\dagger e^{i(\omega_n+\omega_c+\lambda_r) t} +\mathrm{H.c.}\right]\,,\label{eqap:effLM}
\end{align}
where $\alpha=a/b$ represents the $n$-th emitter coupled to sublattice $A/B$, and $\hat{\mathcal{O}}_{r,k}$, $\lambda_r$ and $F^\alpha_{n,r,k}$ are given by
\vspace{10pt}
\renewcommand{\arraystretch}{3.5} 
\setlength{\tabcolsep}{8pt} 

\noindent 
\resizebox{\textwidth}{!}{ 
\begin{tabular}{
    >{\centering\arraybackslash}m{1cm}  
    >{\centering\arraybackslash}m{1cm}  
    >{\centering\arraybackslash}m{2cm}  
    >{\centering\arraybackslash}m{5cm}  
    >{\centering\arraybackslash}m{5cm}
}
\toprule
$r$ & $\hat{\mathcal{O}}_{r,k}$ & $\lambda_r$ & $F^A_{n,r,k}$ & $F^B_{n,r,k}$ \\ 
\hline
1 & $\hat{p}_k$ & \(\lambda(k) + \frac{3\Omega}{2}\) & 
\(\begin{aligned}
    -e^{-i\theta_k}e^{i\beta_k}J_1(z_k) 
    \cos(\frac{\phi_k}{2}) 
\end{aligned}\) & 
\(\begin{aligned}
    -e^{-i\beta_k}J_1(z_k) 
    \cos(\frac{\phi_k}{2}) 
\end{aligned}\) \\
\hline
2 & $\hat{q}_k$ & \(-\lambda(k) + \frac{3\Omega}{2}\) & 
\(\begin{aligned}
    -e^{-i\theta_k}e^{i\beta_k}J_1(z_k) 
    \sin(\frac{\phi_k}{2}) 
\end{aligned}\) & 
\(\begin{aligned}
    -e^{-i\beta_k}J_1(z_k) 
    \sin(\frac{\phi_k}{2}) 
\end{aligned}\) \\
\hline
3 & $\hat{p}_k$ & \(\lambda(k) + \frac{\Omega}{2}\) & 
\(\begin{aligned}
    J_0(z_k) 
    \cos(\frac{\phi_k}{2}) \\
    -i e^{-i\theta_k} e^{i\beta_k} J_1(z_k) 
    \sin(\frac{\phi_k}{2}) 
\end{aligned}\)&
\(\begin{aligned}
    e^{-i \theta_k}J_0(z_k) 
    \cos(\frac{\phi_k}{2}) \\
    +i e^{-i\beta_k} J_1(z_k) 
    \sin(\frac{\phi_k}{2})
\end{aligned}\)\\
\hline
4 & $\hat{q}_k$ & \(-\lambda(k) + \frac{\Omega}{2}\) & 
\(\begin{aligned}
    J_0(z_k) 
    \sin(\frac{\phi_k}{2}) \\
    +ie^{-i\theta_k} e^{i\beta_k} J_1(z_k) 
    \cos(\frac{\phi_k}{2}) 
\end{aligned}\)&
\(\begin{aligned}
    e^{-i\theta_k}J_0(z_k) 
    \sin(\frac{\phi_k}{2}) \\
    -i e^{-i\beta_k} J_1(z_k) 
    \cos(\frac{\phi_k}{2}) 
\end{aligned}\)\\
\hline
5 & $\hat{p}_k$ & \(\lambda(k) - \frac{\Omega}{2}\) &
\(\begin{aligned}
    -iJ_0(z_k) 
    \sin(\frac{\phi_k}{2}) \\
    + e^{-i\theta_k}e^{i\beta_k}J_1(z_k) 
    \cos(\frac{\phi_k}{2})
\end{aligned}\) & 
\(\begin{aligned}
    i e^{-i\theta_k}J_0(z_k) 
    \sin(\frac{\phi_k}{2}) \\
    +e^{-i\beta_k}J_1(z_k) 
    \cos(\frac{\phi_k}{2})
\end{aligned}\)\\
\hline
6 & $\hat{q}_k$ & \(-\lambda(k) - \frac{\Omega}{2}\) & 
\(\begin{aligned}
    iJ_0(z_k) 
    \cos(\frac{\phi_k}{2}) \\
    + e^{-i\theta_k}e^{i\beta_k}J_1(z_k) 
    \sin(\frac{\phi_k}{2})
\end{aligned}\)& 
\(\begin{aligned}
    -i e^{-i\theta_k}J_0(z_k) 
    \cos(\frac{\phi_k}{2}) \\
    +e^{-i\beta_k}J_1(z_k) 
    \sin(\frac{\phi_k}{2})
\end{aligned}\)\\
\hline
7 & $\hat{p}_k$ & \(-\lambda(k) - \frac{3\Omega}{2}\) &
\(\begin{aligned}
    ie^{-i\theta_k}e^{i\beta_k}J_1(z_k) 
    \sin(\frac{\phi_k}{2})
\end{aligned}\)& 
\(\begin{aligned}
    -ie^{-i\beta_k}J_1(z_k) 
    \sin(\frac{\phi_k}{2})
\end{aligned}\)\\
\hline
8 & $\hat{q}_k$ & \(-\lambda(k) - \frac{3\Omega}{2}\) & 
\(\begin{aligned}
    -i e^{-i\theta_k}e^{i\beta_k}J_1(z_k) 
    \cos(\frac{\phi_k}{2})
\end{aligned}\)& 
\(\begin{aligned}
    ie^{-i\beta_k}J_1(z_k) 
    \cos(\frac{\phi_k}{2})
\end{aligned}\) \\
\hline
\end{tabular}
}
\par\vspace{10pt}

We can see from Eq.~\eqref{eqap:effLM}, that the effective detunings of the light-matter couplings are given by $\pm(\omega_n-\omega_c - \lambda_r)$ and $\pm(\omega_n+\omega_c+\lambda_r)$. Since we typically work in regimes where the cavity energy $\omega_c$ is much larger than the hopping amplitudes $J$ and $J'$ and we focus on the intermediate-frequency regime $\Omega\in2\omega(k)\sim J$, this means that $\lambda_r \ll \omega_c$. In that case, the detunings reduce to $\pm(\omega_n-\omega_c)$ and $\pm(\omega_n+\omega_c)$. Besides, in the weak coupling approximation, we also consider, i.e., $g\ll \omega_n,\omega_c$, that allows us to neglect the rapidly oscillating terms and only retain the particle number conserved terms $\hat{\sigma}_n^\dagger \hat{\mathcal{O}}_{r,k}$ and $\hat{\sigma}_n \hat{\mathcal{O}}_{r,k}^\dagger$ by RWA. Therefore, the effective Hamiltonian in the interaction picture can be simplified to:
\begin{align}
    \hat{H}_\mathrm{eff}(t) &= \frac{g}{\sqrt{2N}}\sum_{n,r,k} \left[F^\alpha_{n,r,k}e^{ikj_n} \hat{\sigma}_n^\dagger  \hat{\mathcal{O}}_{r,k} e^{i(\omega_n-\omega_c-\lambda_r) t} +\mathrm{H.c.}\right]\,,\label{eqap:effLM2}
\end{align}
which is the one we will consider for the calculations.

\section{Approximating the quasi-energy spectrum of the time-dependent bath Hamiltonians~\label{app:quasi}}

In Eq.~\eqref{eqap:effH}, we obtain the eigenvalues $\omega_c\pm\lambda(k)$ and eigenstate $\hat{p}_k^\dagger[\hat{q}_k^\dagger]\ket{\mathrm{vac}}_B$ of the effective time-independent bath Hamiltonian after performing several transformations and approximations. Undoing the transformations we use to derive it, we can obtain the solutions in the original frame, and thus obtain the approximated Floquet states and their corresponding quasi-energies. For example, the Floquet states can be obtained by:
\begin{align}
\ket{\Psi_\pm(t)}_B&=\hat{R}_d(t)\hat{\Lambda}_1 \hat{R}_\Omega(t)\hat{\Lambda}_2 e^{- i(\omega_c\pm \lambda(k))t}\hat{p}_k^\dagger[\hat{q}_k^\dagger]\ket{\mathrm{vac}}_B \\ \notag
&=e^{-i\varepsilon_{\pm} t} \left( \hat{R}_d(t) \hat{\Lambda}_1 \hat{R}_\Omega(t) \hat{\Lambda}_2 e^{i\left(\varepsilon_\pm-\omega_c\mp \lambda(k)\right)t}\hat{p}_k^\dagger[\hat{q}_k^\dagger]\ket{\mathrm{vac}}_B\right) =e^{-i\varepsilon_\pm t} \ket{\Phi_\pm(t)}\,.
\end{align}

Notice that we have $\hat{R}_d(t+T)=\hat{R}_d(t)$ and $\hat{R}_\Omega(t+T)=-\hat{R}_\Omega(t)$. Regarding the quasi-energies, if we consider the periodicity of the Floquet state $\ket{\Phi_\pm(t)}=\ket{\Phi_\pm(t+T)}$, this requires $\left(\varepsilon_\pm-\omega_c\mp \lambda(k)\right)T=(2m+1)\pi$ with $m\in \mathbb{Z}$. Thus, the relation between the effective time-independent eigenvalues, $\omega_c\pm\lambda(k)$, and the real quasi-energies is:
\begin{align}
\varepsilon_{\pm} = \omega_c +\frac{2m+1}{2}\Omega \pm \lambda(k), \quad m\in \mathbb{Z}.
\end{align}

\section{Derivation of the effective master equation for time-dependent baths~\label{app:meq}}

In this section, we aim at deriving an effective description of the emitters coupled to the time-dependent bath. For that, we will use the approximated effective light-matter coupling Hamiltonians of Eqs.~\eqref{eqap:effLM2}. Considering emitters with detunings $\Delta_n=\omega_n-\omega_c$, this light-matter Hamiltonian in the interaction picture can be written as:
\begin{align}
    \hat{H}_\mathrm{eff}(t) &=\sum_ne^{i\Delta_n t} \hat{\sigma}_n^\dagger \hat{B}_n(t) +\mathrm{H.c.}\,,
\end{align}
where $\hat{B}_n(t)=\frac{g}{\sqrt{2N}}\sum_{r,k} F^\alpha_{n,r,k}e^{ikj_n}\mathcal{O}_{r,k} e^{-i\lambda_r t}$ is the bath operator.

In weak coupling limit, we can eliminate the bath degrees of freedom and get the reduced master equation in the Born-markov approximation for the emitters' dynamics only:
\begin{align}
    \dot{\hat{\rho}}_S = -\int_0^\infty ds\, \text{tr}_B\left\{ \left[\hat{H}_\mathrm{eff}(t),\, \left[\hat{H}_\mathrm{eff}(t-s),\, \hat{\rho}_S\hat{\rho}_B\right]\right] \right\}\, ,
\end{align}

Considering a vacuum waveguide, i.e., $\langle \hat{\alpha}_k \hat{\beta}_{k'}^\dagger\rangle=\delta_{\alpha\beta} \delta_{kk'}$ and $\langle \hat{\alpha}_k^\dagger \hat{\beta}_{k'}\rangle=\langle \hat{\alpha}_k^\dagger \hat{\beta}_{k'}^\dagger\rangle=\langle \hat{\alpha}_k \hat{\beta}_{k'}\rangle=0$ where $\alpha ( \beta)=p,q$, we get a simplified master equation
\begin{align}
        \dot{\hat{\rho}}_S &=\sum_{n,m} \left[\hat{\sigma}_m \hat{\rho}_S,\, \hat{\sigma}_n^\dagger\right] e^{i(\Delta_n-\Delta_m)t}\int_0^\infty ds\, e^{i\Delta_m s}C_{nm}(t,t-s) \\ \notag
        &+\sum_{n,m} \left[  \hat{\sigma}_m,\,\hat{\rho}_S\hat{\sigma}_n^\dagger\right] e^{i(\Delta_n-\Delta_m)t}\int_0^\infty ds\, e^{-i\Delta_n s}K_{nm}(t,t-s)\,,
\end{align}
where we define the time correlators:
\begin{align}
    &C_{nm}(t,t-s)=\langle \hat{B}_{n}(t)\hat{B}^\dagger_{m}(t-s)\rangle=\frac{g^2}{2N}\sum_k e^{ikj_{nm}}\sum_{r,r'} F^\alpha_{n,r,k} F_{m,r',k}^{\beta,*} e^{-i(\lambda_r-\lambda_{r'})t} e^{-i\lambda_{r'}s}\,, \\ \notag
    &K_{nm}(t,t-s)=
    \langle \hat{B}_{n}(t-s)\hat{B}^\dagger_{m}(t)\rangle=\frac{g^2}{2N}\sum_k e^{ikj_{nm}}\sum_{r,r'} F^\alpha_{n,r,k} F_{m,r',k}^{\beta,*} e^{-i(\lambda_r-\lambda_{r'})t} e^{i\lambda_{r}s}\,,
\end{align}

Thus, to obtain the master equation we have to solve the following integrals:
\begin{align}
    e^{i(\Delta_n-\Delta_m)t}\int_0^\infty ds\, e^{i\Delta_m s}C_{nm}(t,t-s) = \frac{g^2}{2N}\sum_k e^{ikj_{nm}}\sum_{r,r'} F^\alpha_{n,r,k} F_{m,r',k}^{\beta,*} e^{i(\Delta_n-\Delta_m-\lambda_r+\lambda_{r'})t} \frac{i}{\Delta_m+i0^+-\lambda_{r'}}\,,\\ \notag
    e^{i(\Delta_n-\Delta_m)t}\int_0^\infty ds\, e^{-i\Delta_n s}K_{nm}(t,t-s) = \frac{g^2}{2N}\sum_k e^{ikj_{nm}}\sum_{r,r'} F^\alpha_{n,r,k} F_{m,r',k}^{\beta,*} e^{i(\Delta_n-\Delta_m-\lambda_r+\lambda_{r'})t} \frac{-i}{\Delta_n-i0^+-\lambda_{r}}\,,\\ \notag
\end{align}

Let us now distinguish between three different situations:

\uppercase\expandafter{\romannumeral1}. The case of single emitter with detuning $\Delta$ and coupled to sublattice $\alpha$: when $r\neq r'$, the terms $e^{i(\Delta_n-\Delta_m-\lambda_r+\lambda_{r'})t}$ oscillate with multiples of the drive frequency $\Omega$ and can be eliminated by RWA, and the integrals thus can be simplified as:
\begin{align} \label{eq:derive_selsenergy}
    e^{i(\Delta_n-\Delta_m)t}\int_0^\infty ds\, e^{i\Delta_m s}C_{nm}(t,t-s) &=i \frac{g^2}{2N}\sum_{r,k}  \frac{|F^{\alpha}_{n,r,k}|^2 }{\Delta+i0^+-\lambda_{r}} =i\Sigma_{\mathrm{eff}}(\Delta+i0^+) = i(\delta \omega - i\Gamma/2)\,, \\ \notag
    e^{i(\Delta_n-\Delta_m)t}\int_0^\infty ds\, e^{-i\Delta_n s}K_{nm}(t,t-s) 
    &= -i\frac{g^2}{2N}\sum_{r,k}   \frac{|F^{\alpha}_{n,r,k}|^2}{\Delta-i0^+-\lambda_{r}} =-i\Sigma_{\mathrm{eff}}(\Delta-i0^+) = -i(\delta \omega + i\Gamma/2)\,,
\end{align}
where $\Sigma_{\mathrm{eff}}(\Delta+i0^+)$ is the self-energy for single emitter.
The master equation in the interaction picture is given by
\begin{align}
    \dot{\hat{\rho}}_S=-i\left[\delta\omega \hat{\sigma}^\dagger \hat{\sigma},\, \hat{\rho}_S\right] + \Gamma \mathcal{D}[\hat{\sigma}, \hat{\sigma}^\dagger]\hat{\rho}_S\,,
\end{align}
where $\mathcal{D}[\hat{a}, \hat{b}]\hat{\rho}=\hat{a}\hat{\rho}\hat{b} - (\hat{b}\hat{a}\hat{\rho} + \hat{\rho}\hat{b}\hat{a})/2$ is the standard Lindblad term.

\uppercase\expandafter{\romannumeral2}. The case of two identical emitters with $\Delta_n=\Delta_m=\Delta$ coupled to sublattice $\alpha$ and $\beta$: when $r\neq r'$, the terms $e^{i(\Delta_n-\Delta_m-\lambda_r+\lambda_{r'})t}$ oscillate with multiples of the drive frequency $\Omega$ and can be eliminated by RWA, and the integrals thus can be simplified to:
\begin{align}
    e^{i(\Delta_n-\Delta_m)t}\int_0^\infty ds\, e^{i\Delta_m s}C_{nm}(t,t-s) 
    &= i\frac{g^2}{2N}  \sum_{r,k}  \frac{F^{\alpha}_{n,r,k} F_{m,r,k}^{\beta,*} }{\Delta+i0^+-\lambda_{r}} e^{ikj_{nm}}\\ \notag
    &=i\Sigma^{\alpha\beta}_{nm}(\Delta+i0^+)=i(G^{\alpha\beta}_{nm}-i\Gamma^{\alpha\beta}_{nm}/2) \\ \notag
    e^{i(\Delta_n-\Delta_m)t}\int_0^\infty ds\, e^{-i\Delta_n s}K_{nm}(t,t-s) &=-i\frac{g^2}{2N}  \sum_{r,k}  \frac{F^{\alpha}_{n,r,k} F_{m,r,k}^{\beta,*} }{\Delta-i0^+-\lambda_{r}} e^{ikj_{nm}} \\ \notag
    &=-i\Sigma^{\alpha\beta}_{nm}(\Delta-i0^+)=-i(G^{\alpha\beta}_{nm}+i\Gamma^{\alpha\beta}_{nm}/2)\,,
\end{align}

The master equation in the interaction picture is given by:
\begin{align}
    \dot{\hat{\rho}}_S=-i\left[\sum_{n,m} G^{\alpha\beta}_{nm} \hat{\sigma}_n^\dagger \hat{\sigma}_m,\, \hat{\rho}_S\right] + \sum_{n,m}\Gamma^{\alpha\beta}_{nm} \mathcal{D}[\hat{\sigma}_m, \hat{\sigma}_n^\dagger]\hat{\rho}_S\,,
\end{align}

When we set the frequency $\Delta$ to 0 or $\pi$-gap, the decay process is prohibited $\Gamma_{nm}=0$, so the dynamics are dominated by the dipole-dipole interaction, described as
\begin{align}
    \hat{H}_\mathrm{QE} = \sum_{n,m} G^{\alpha\beta}_{nm} \hat{\sigma}_n^\dagger \hat{\sigma}_m\,.
\end{align}

\uppercase\expandafter{\romannumeral3}. The case of emitter with frequency difference $|\Delta_1-\Delta_2|=\Omega$ and coupled to sublattice $\alpha$ and $\beta$: Two of terms $e^{i(\Delta_n-\Delta_m-\lambda_r+\lambda_{r'})t}$ are retained, while the others oscillate with multiples of the drive frequency $\Omega$ and are therefore eliminated by RWA. The first is $n=m$, $r=r'$, which corresponds to the self-energy of a single emitter in Eq.~\eqref{eq:derive_selsenergy}, and the second is $n\neq m$ and $|r-r'|=2$, which is the interaction mediated by the driving photon that compensates for the emitter frequency difference. For $n\neq m$, we get
\begin{align}
    &e^{i(\Delta_n-\Delta_m)t}\int_0^\infty ds\, e^{i\Delta_m s}C_{nm}(t,t-s) \\ \notag
    &=  ie^{i(\Delta_n-\Delta_m+\Omega)t}\frac{g^2}{2N}\sum_{k,r=1}^{r=6}     \frac{F^{\alpha}_{n,r+2,k} F_{m,r,k}^{\beta,*}}{\Delta_m+i0^+-\lambda_r} e^{ikj_{nm}} + ie^{i(\Delta_n-\Delta_m-\Omega)t}\frac{g^2}{2N}\sum_{k,r=1}^{r=6}   \frac{F^{\alpha}_{n,r,k} F_{m,r+2,k}^{\beta,*}}{\Delta_m+i0^+-\lambda_{r+2}} e^{ikj_{nm}}\,,\\ \notag
    &e^{i(\Delta_n-\Delta_m)t}\int_0^\infty ds\, e^{-i\Delta_n s}K_{nm}(t,t-s)\,, \\ \notag
    &=  -ie^{i(\Delta_n-\Delta_m+\Omega)t}\frac{g^2}{2N}\sum_{k,r=1}^{r=6}     \frac{F^{\alpha}_{n,r+2,k} F_{m,r,k}^{\beta,*}}{\Delta_n-i0^+-\lambda_{r+2}} e^{ikj_{nm}} - ie^{i(\Delta_n-\Delta_m-\Omega)t}\frac{g^2}{2N}\sum_{k,r=1}^{r=6}   \frac{F^{\alpha}_{n,r,k} F_{m,r+2,k}^{\beta,*}}{\Delta_n-i0^+-\lambda_{r}}e^{ikj_{nm}}\,.
\end{align}

When we set the frequency $\Delta_n$ within $0$- or $\pi$-gap (assume $\Delta_1-\Delta_2=\Omega$), the decay process is prohibited. We can use RWA to eliminate the term in the integrals oscillating with frequency $2\Omega$ and get the Hamiltonian:
\begin{align}
    \hat{H}_\mathrm{QE} =  \delta \omega_1 \hat{\sigma}_1^\dagger \hat{\sigma}_1 + \delta \omega_2 \hat{\sigma}_2^\dagger \hat{\sigma}_2 + G_{12}^{\alpha\beta,\Omega}\left(\hat{\sigma}_1^\dagger \hat{\sigma}_2 e^{i(\Delta_1-\Delta_2-\Omega)t} + \mathrm{H.c.}\right)\,,
\end{align}
where $\delta\omega_n=\delta\omega(\Delta_n)=\mathrm{Re}[\Sigma_\mathrm{eff}(\Delta_n+i0^+)]$ is Lamb shift, since the frequencies are different, and the coupling strength $G_{12}^{\alpha\beta,\Omega}$ is given
by:
\begin{align} 
G_{12}^{\alpha\beta,\Omega}&=\frac{g^2}{2N}\sum_{k,r=1}^{r=6}    \frac{F^{\alpha}_{1,r,k} F_{2,r+2,k}^{\beta,*}}{\Delta_2-\lambda_{r+2}} e^{ikj_{12}}\,,
\end{align}

A crucial point is that the analytical solution quantitatively fails under specific conditions. Near the $\pi$-gap, the inequality $|\gamma_k|\ll \Omega,|2\omega_d(k)|$ and $|2\omega_d(k) + \Omega| \gg |2\omega_d(k) - \Omega|$ are always satisfied, ensuring the RWA's validity. For the 0-gap case, in the absence of emitters or when emitters share identical frequencies ($\Delta_n = \Delta$), both the rotating and counter-rotating terms remain negligible since they are off-resonate. Consequently, all characteristics near the 0-gap, such as the band structure, self-energy, and emitter dynamics, are accurately described by the static SSH Hamiltonian. Neglecting either or both terms under these conditions yields correct results. However, the situation changes fundamentally when the emitter frequency difference is resonant with the drive frequency, $|\Delta_1 - \Delta_2| = \Omega$ in 0-gap. First, the static Hamiltonian alone induces neither decay nor oscillation in the emitters. Second, a single driving photon can mediate the inter-emitter interaction by precisely compensating for their frequency difference. Third, this resonance renders counter-rotating relevant; they can no longer be neglected as in the off-resonant case. The problem arises because near the 0-gap, under this resonance condition, the contributions from the rotating and counter-rotating terms are significant and comparable in magnitude. Consequently, neglecting the counter-rotating term becomes invalid, leading to the frequency shift mismatch between theory and numerical results observed in Fig.~\ref{fig:exchange_Omega}(a).

\end{widetext}
\bibliographystyle{apsrev4-2}
\bibliography{references}

\end{document}